\documentclass[twocolumn,tighten,times,twocolappendix]{aastex631}
\usepackage{amsmath}
\usepackage{multirow}
\usepackage{hyperref}
\usepackage{scalerel}
\usepackage{footnote}
\usepackage{rotating}
\usepackage{makecell}
\usepackage{threeparttable}

\DeclareRobustCommand{\ion}[2]{%
\relax\ifmmode
\ifx\testbx\f@series
{\mathbf{#1\,\mathsc{#2}}}\else
{\mathrm{#1\,\mathsc{#2}}}\fi
\else\textup{#1\,{\mdseries\textsc{#2}}}%
\fi}

\def\arcsec{$''$}
\def\arcmin{$'$}
\def\HII{\ion{H}{II}}
\def\CII{[\ion{C}{II}]}
\def\OI{[\ion{O}{I}]}
\def\Av{$A_{\mathrm{V}}$}
\def\Zprime{$Z^{\prime}_{x}$}
\def\Zx{$Z_{x}$}
\def\Znorm{$\Tilde{Z}^{\prime}_{x}$}
\def\Bfield{$\hat{B}_{\perp}$}
\def\Halpha{H$\alpha$}
\def\sigmaWN{$\sigma_{Z^{\prime}_{WN}}$}
\def\sigmaZx{$\sigma_{Z_{x}}$}
\def\sigmaB{$\sigma_{\hat{B}_{\perp}}$}

\begin{document}

\author[0000-0001-7505-5223]{Akanksha Bij}
\affiliation{Department of Physics, Engineering Physics and Astronomy, Queen’s University, Kingston, ON K7L 3N6, Canada} 

\author[0000-0002-4666-609X]{Laura M. Fissel}
\affiliation{Department of Physics, Engineering Physics and Astronomy, Queen’s University, Kingston, ON K7L 3N6, Canada} 

\author[0000-0002-0915-4853]{Lars Bonne}
\affiliation{SOFIA Science Center, NASA Ames Research Center, Moffett Field, CA 94045, USA} 

\author[0000-0003-3485-6678]{Nicola Schneider}
\affiliation{I. Physik. Institut, University of Cologne, Z\"{u}lpicher Str. 77, D-50937 Cologne, Germany} 

\author{Marc Berthoud}
\affiliation{Center for Interdisciplinary Exploration and Research in Astrophysics (CIERA), 1800 Sherman Avenue, Evanston, IL 60201, USA}
\affiliation{Engineering + Technical Support Group, University of Chicago, Chicago, IL 60637, USA}

\author[0000-0002-3455-1826]{Dennis Lee}
\affiliation{Center for Interdisciplinary Exploration and Research in Astrophysics (CIERA), 1800 Sherman Avenue, Evanston, IL 60201, USA}
\affiliation{Department of Physics \& Astronomy, Northwestern University, 2145 Sheridan Road, Evanston, IL 60208, USA} 

\author[0000-0003-1288-2656]{Giles A. Novak}
\affiliation{Center for Interdisciplinary Exploration and Research in Astrophysics (CIERA), 1800 Sherman Avenue, Evanston, IL 60201, USA}
\affiliation{Department of Physics \& Astronomy, Northwestern University, 2145 Sheridan Road, Evanston, IL 60208, USA}

\author[0000-0001-7474-6874]{Sarah I. Sadavoy}
\affiliation{Department of Physics, Engineering Physics and Astronomy, Queen’s University, Kingston, ON K7L 3N6, Canada} 

\author[0000-0003-2133-4862]{Thushara G. S. Pillai}
\affiliation{MIT Haystack Observatory, 99 Millstone Road , Westford, MA, 01827} 

\author[0000-0001-7020-6176]{Maria Cunningham}
\affiliation{School of Physics, University of New South Wales, Sydney NSW 2052, Australia} 

\author[0000-0001-9429-9135]{Paul Jones}
\affiliation{School of Physics, University of New South Wales, Sydney NSW 2052, Australia} 

\author[0000-0003-2555-4408]{Robert Simon}
\affiliation{I. Physik. Institut, University of Cologne, Z\"{u}lpicher Str. 77, D-50937 Cologne, Germany} 

\correspondingauthor{Akanksha Bij}
\email{a.bij@queensu.ca}

\title{Magnetic Field Alignment Relative to Multiple Tracers in the High-mass Star-forming Region RCW 36}

\begin{abstract}
We use polarization data from SOFIA HAWC+ to investigate the interplay between magnetic fields and stellar feedback in altering gas dynamics within the high-mass star-forming region RCW 36, located in Vela C. This region is of particular interest as it has a bipolar \HII\ region powered by a massive star cluster which may be impacting the surrounding magnetic field. To determine if this is the case, we apply the Histogram of Relative Orientations (HRO) method to quantify the relative alignment between the inferred magnetic field and elongated structures observed in several datasets such as dust emission, column density, temperature, and spectral line intensity maps. The HRO results indicate a bimodal alignment trend, where structures observed with dense gas tracers show a statistically significant preference for perpendicular alignment relative to the magnetic field, while structures probed by photo-dissociation region (PDR) tracers tend to align preferentially parallel relative to the magnetic field. Moreover, the dense gas and PDR associated structures are found to be kinematically distinct such that a bimodal alignment trend is also observed as a function of line-of-sight velocity. This suggests that the magnetic field may have been dynamically important and set a preferred direction of gas flow at the time that RCW 36 formed, resulting in a dense ridge developing perpendicular to the magnetic field. However on filament-scales near the PDR region, feedback may be energetically dominating the magnetic field, warping its geometry and the associated flux-frozen gas structures, causing the observed the preference for parallel relative alignment.
\end{abstract}


\section{Introduction}

Observations and simulations suggest that star formation occurs when density fluctuations undergo gravitational collapse in molecular clouds \citep{2007ARA&A..45..565M}. The interstellar magnetic field is thought to influence the structure and evolution of these molecular clouds through regulating the rate and efficiency at which gas is converted into pre-stellar structures by providing support against collapse and/or directing gas flow \citep{2023ASPC..534..193P}. In the vicinity of massive stars, stellar feedback in the form of winds, outflows, and radiation pressure can further alter the dynamical and chemical evolution of the molecular cloud \citep[e.g.,][]{1979MNRAS.186...59W, 2009ApJ...694L..26G, 2011ApJ...731...91L, 2019ApJ...872..187C}. 

Stellar feedback has also been observed to reshape magnetic field geometries around expanding ionized bubbles \citep[e.g.,][]{1989ApJ...336..808H, 2018A&A...609L...3S, 2019A&A...632A..68T}, which is consistent with predictions from magnetohydrodynamic (MHD) simulations \citep[e.g.][]{2007ApJ...671..518K}. However the combined impact of both magnetic fields and stellar feedback in high-mass star forming regions remains poorly understood due to various constraints. For instance, simulating the effect of stellar feedback on the parent molecular cloud requires complex sub-grid physics and  demanding computational resources \citep[e.g.,][]{2014MNRAS.442..694D, 2020MNRAS.492..915G}. Furthermore measuring the magnetic field strength through observations is challenging. 
While numerous observational techniques such as Zeeman splitting \citep{2012ARA&A..50...29C, 2019FrASS...6...66C}, Faraday rotation \citep{2018A&A...614A.100T} and the Davis-Chandrasekhar-Fermi (DCF) method \citep{1951PhRv...81..890D, 1953ApJ...118..113C} applied to polarized light have been used in the past \citep[e.g.,][]{2006Sci...313..812G, 2015ApJ...799...74P, 2016A&A...586A.138P}, each technique has limitations and/or only provides partial information about the magnetic field structure. 

Of all methods to study magnetic fields, polarized dust emission is the most commonly used observational tracer in dense molecular clouds. The plane-of-sky magnetic field orientation can be inferred from the linearly polarized emission of non-spherical dust grains, which are thought to align their long axes perpendicular to the local magnetic field lines on average \citep{1951ApJ...114..206D, 2015ARA&A..53..501A}. Dust polarization angle maps can therefore be used as a proxy for the magnetic field orientation weighted by density, dust grain efficiency, and dust opacity. Various comparisons between the orientation of the magnetic field lines to the orientation of molecular cloud structures have been studied to gain insight into the role of the magnetic field in the star-forming process \citep[e.g.,][]{2008ApJ...680..428G, 2009MNRAS.399.1681T, 2013MNRAS.436.3707L, 2016A&A...586A.138P, 2020NatAs...4.1195P}.

A numerical method known as the Histogram of Relative Orientations (HRO) was developed by \cite{2013ApJ...774..128S} to statistically characterize this comparison by measuring the relative alignment between the magnetic field orientation, and the orientation of iso-contours of elongated structures measured from a gradient field. Several HRO studies have found that the relative alignment between interstellar structures and the plane-of-sky magnetic field orientation is dependent on density and column density \citep[e.g.,][]{2017A&A...607A...2S, 2020NatAs...4.1195P, 2020MNRAS.497.4196S}. Most notably,  \cite{2016A&A...586A.138P} implemented the HRO method for 10 nearby ($<$ 400 pc) molecular clouds using polarimetry data with a resolution of 10\arcmin{} at 353 GHz from the \textit{Planck} satellite. They found that the overall alignment of elongated structures transitioned from either random or preferentially parallel relative to the magnetic field at lower column densities, to preferentially perpendicular at higher column densities, with the switch occurring at different critical column densities for each cloud. In simulations this signature transition to perpendicular alignment for an increasing column density has been seen for strong magnetic fields that are significant relative to turbulence and able to influence the gas dynamics (i.e. dynamically important)
 \citep[e.g.,][]{2017A&A...607A...2S, 2020MNRAS.499.4785K}. 

The HRO method has since been applied to younger and more distant giant molecular clouds, such as Vela C (distance of $\sim$ 900 pc), whose magnetic field morphology was inferred by the BLASTPol (Balloon-borne Large-Aperture Submillimetre Telescope for Polarimetry) instrument at 250, 350 and 500 $\mu$m \citep{2016ApJ...824..134F}. The BLASTPol-inferred magnetic field was compared to column densities derived from \emph{Herschel} observations by \cite{2017A&A...603A..64S}, and to the integrated intensities of molecular gas tracers by \cite{2019ApJ...878..110F}. Both studies found a similar tendency for elongated structures to align preferentially parallel relative to the magnetic field for low column densities or low-density gas tracers, which then switched to preferentially perpendicular for higher column densities or high-density gas tracers. However with a full width at half maximum (FWHM) resolution of $\sim$3\arcmin{}, the BLASTPol observations are only able to probe the Vela C magnetic field geometry on cloud-scales ($>$ 1 pc).

In this work, we extend these HRO studies to filament-scales ($\sim$0.1\textendash1 pc) by using higher resolution polarimetry data from the Stratospheric Observatory for Infrared Astronomy (SOFIA) High-resolution Airborne Wide-band Camera (HAWC+) instrument at 89 $\mu$m (Band C) and 214 $\mu$m (Band E), with angular resolution of 7.8\arcsec{} (0.03 pc) and 18.2\arcsec{} (0.08 pc), respectively \citep{2018JAI.....740008H}. Moreover, we focus on the role of magnetic fields in high-mass star formation by targeting the densest region within Vela C, known as RCW 36 (visual extinction of \Av\ $>$ 100 mag) which is within a parsec of a ionizing young ($\sim$ 1 Myr) OB cluster responsible for powering a bipolar \HII\ nebula within the region \citep{2013A&A...550A..50M, 2013A&A...558A.102E}.  

Previous HRO studies have mostly compared the relative orientation of the magnetic field to molecular gas structures. Since this study aims to understand the role of stellar feedback, we wish to additionally apply the HRO analysis to structures associated with the photodissociation region (PDR). The PDR is the interfacing boundary between the ionized \HII\ region and surrounding molecular cloud where far-ultraviolet (FUV) photons with energies in the range of 6\textendash13.6 eV dominate and dissociate H$_{2}$ and CO molecules \citep{1997ARA&A..35..179H}. Several recent \HII\ region studies have used observations of \CII\ since it traces the PDR \citep[e.g.,][]{2018A&A...617A..45S, 2019ApJ...882...11A, 2019Natur.565..618P, 2021SciA....7.9511L, 2022A&A...659A..77B,  2023ApJ...946....8T}. For RCW 36, the bipolar \HII\ region was investigated by \cite{2022ApJ...935..171B} who examined the kinematics of \CII\ 158 $\mu$m and \OI\ 63 $\mu$m data from the SOFIA legacy project FEEDBACK \citep{2020PASP..132j4301S}. 

We build upon these studies by applying the HRO method to multiple complementary observations tracing column density, temperature, molecular gas, as well as the PDR, to construct a more complete picture of how the magnetic field is affecting star formation within RCW 36. The paper is organized as follows: Section \ref{sec:Data} describes the observations. Section \ref{sec:Structure} details the physical structure of the RCW 36 and its magnetic field morphology, including noteworthy regions. Section \ref{sec:Methods} discusses the HRO method, with the results presented in Section \ref{Sec:Results}. In Section \ref{sec:Discussion}, we interpret the results and compare this work to other studies. Finally in Section \ref{sec:Conclusion}, we summarize our main conclusions.

\section{Data}
\label{sec:Data}

\begin{table*}
\caption{Summary of all datasets used for Single Map HRO analysis described in section \ref{sec:Methods_single}}
\label{tab:Obs}
\centering
    \begin{threeparttable}
    \begin{tabular}{cccccc}
        \hline
        Instrument & Data Type & $\theta_{beam}$  & Pixel size & $\theta_{G}\,^{a}$ & l$_{G}\,^{b}$  \\
        & & (arcsec) & (arcsec) & (arcsec) & (pixels)  \\
        \hline \hline      
        \multirow{2}{*}{SOFIA/HAWC+} & 
        89 $\mu$m IQU & 7.8 & 1.9 & 5.8 & 3.0 \\ 
        & 214 $\mu$m IQU & 18.2 & 4.6 & 6.1 & 3.1 \\
        \hline
        
        \multirow{2}{*}{\emph{Herschel}/PACS} &
         70 $\mu$m &  7.5 & 3.2 & 5.8 & 3.0 \\
         & 160 $\mu$m & 13.4 & 6.4 & 5.8 & 3.0 \\
        \hline
        
        \multirow{3}{*}{\emph{Herschel}/SPIRE} & 
        250 $\mu$m & 18 & 6.0 & 6.0 & 3.1 \\
        & 350 $\mu$m & 25 & 10.0 & 8.3 & 4.3 \\
        & 500 $\mu$m & 36 & 14.0 & 12.0 & 6.2 \\
        \hline

        \multirow{2}{*}{\emph{Herschel}-derived} & 
        Column Density N(H$_2$) & 18 & 3.0 & 6.0 & 3.1 \\
        & Temperature & 36 & 3.0 & 12.0 & 6.2 \\
        \hline
        
        \multirow{2}{*}{\emph{Spitzer}/IRAC} & 
        3.6 $\mu$m & 1.7 & 0.6 & 1.8 & 3.0 \\
        & 4.5 $\mu$m & 1.7 & 0.6 & 1.8 & 3.0 \\
        \hline
        
        ALMA ACA & 1.1--1.4 mm & 5.4 & 0.9 & 2.8 & 3.0 \\
        \hline
        ALMA 12-m & 1.1--1.4 mm & 1.4 & 0.2 & 0.7 & 3.0 \\
        \hline

        \multirow{2}{*}{APEX/LAsMA} & 
        $^{12}$CO (3--2) & 18.2 & 9.6 & 6.1 & 3.1 \\
        & $^{13}$CO (3--2) & 18.2 & 10.0 & 6.1 & 3.1 \\
        \hline
        
        \multirow{2}{*}{SOFIA/upGREAT} &
         \CII\ $^{3}$P$_{3/2} \rightarrow ^{3}$P$_{1/2}$ & 20 & 3.5 & 6.7 & 3.4 \\
         & \OI\ $^{3}$P$_{1} \rightarrow ^{3}$P$_{2}$ & 30 & 15.0 & 10.0 & 5.1 \\
        \hline
        
        \multirow{2}{*}{Mopra} &
         HNC (1--0) & 36 & 12.0 & 12.0 & 6.2 \\
         & N$_{2}$H$^{+}$ (1--0) & 36 & 12.0 & 12.0 & 6.2 \\
         & C$^{18}$O (1--0) & 33 & 12.0 & 11.0 & 5.6 \\
         \hline
        
         \hline
    \end{tabular}
    \begin{tablenotes}
        \small
        \item $^{a}$ The FWHM angular size of the Gaussian derivative kernel $G(x,y)$ used in the HRO analysis with HAWC+ Band C from Eq. \ref{Eqn:Gauss_smooth} (after projection onto the Band C grid, if applicable). 
        \item $^{b}$ The size of $G(x,y)$ from $^{a}$ in pixels
    \end{tablenotes}
    \end{threeparttable}
\end{table*}

\begin{figure*}
    \centering  \includegraphics{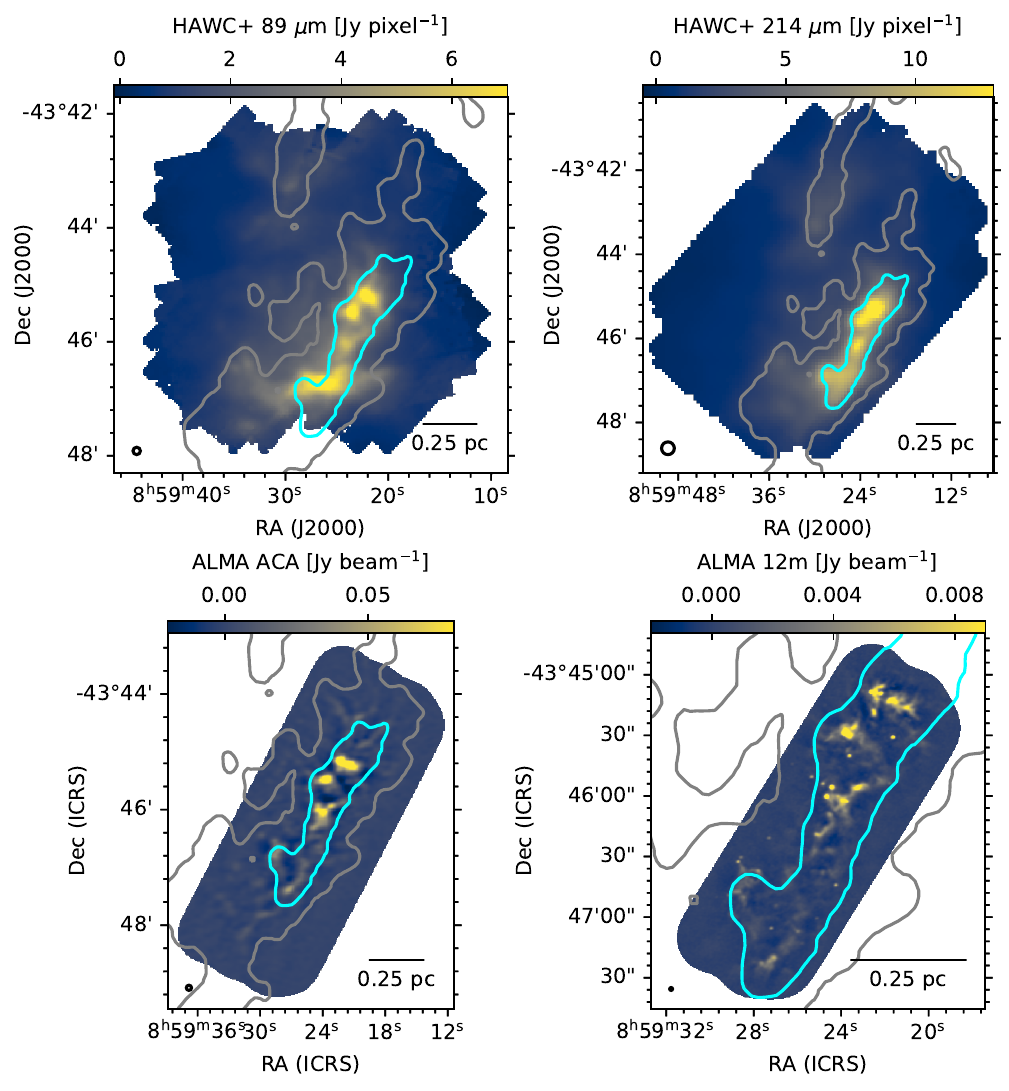}
    \caption{Far-infrared and millimeter intensity maps of the RCW 36 region presented for the first time in this work. \textit{Top Left:} SOFIA HAWC+ 89 $\mu$m (Band C) total intensity in Jy pixel$^{-1}$. \textit{Top Right:} SOFIA HAWC+ 214 $\mu$m (Band E)  total intensity in Jy pixel$^{-1}$. \textit{Bottom Left:} ALMA ACA 1.1--1.4 mm (Band 6) continuum in Jy beam$^{-1}$ \textit{Bottom Right:} ALMA 12m 1.1--1.4 mm (Band 6) continuum in Jy beam$^{-1}$. All color maps are in linear scale. A 0.25 pc scale bar is shown on the bottom right. The FWHM beam size is shown on bottom left (beam and pixel sizes are given in Table \ref{tab:Obs}). The gray and cyan contours show \emph{Herschel}-derived column densities at values of 1.5$\times 10^{22}$ and 4.7$\times 10^{22}$  cm$^{-2}$, respectively. These contours are shown as a position reference on all maps of RCW 36 presented in this work.} 
    \label{fig:NewData}  
\end{figure*}

\begin{figure*}
    \centering  \includegraphics[width=0.80\textwidth]{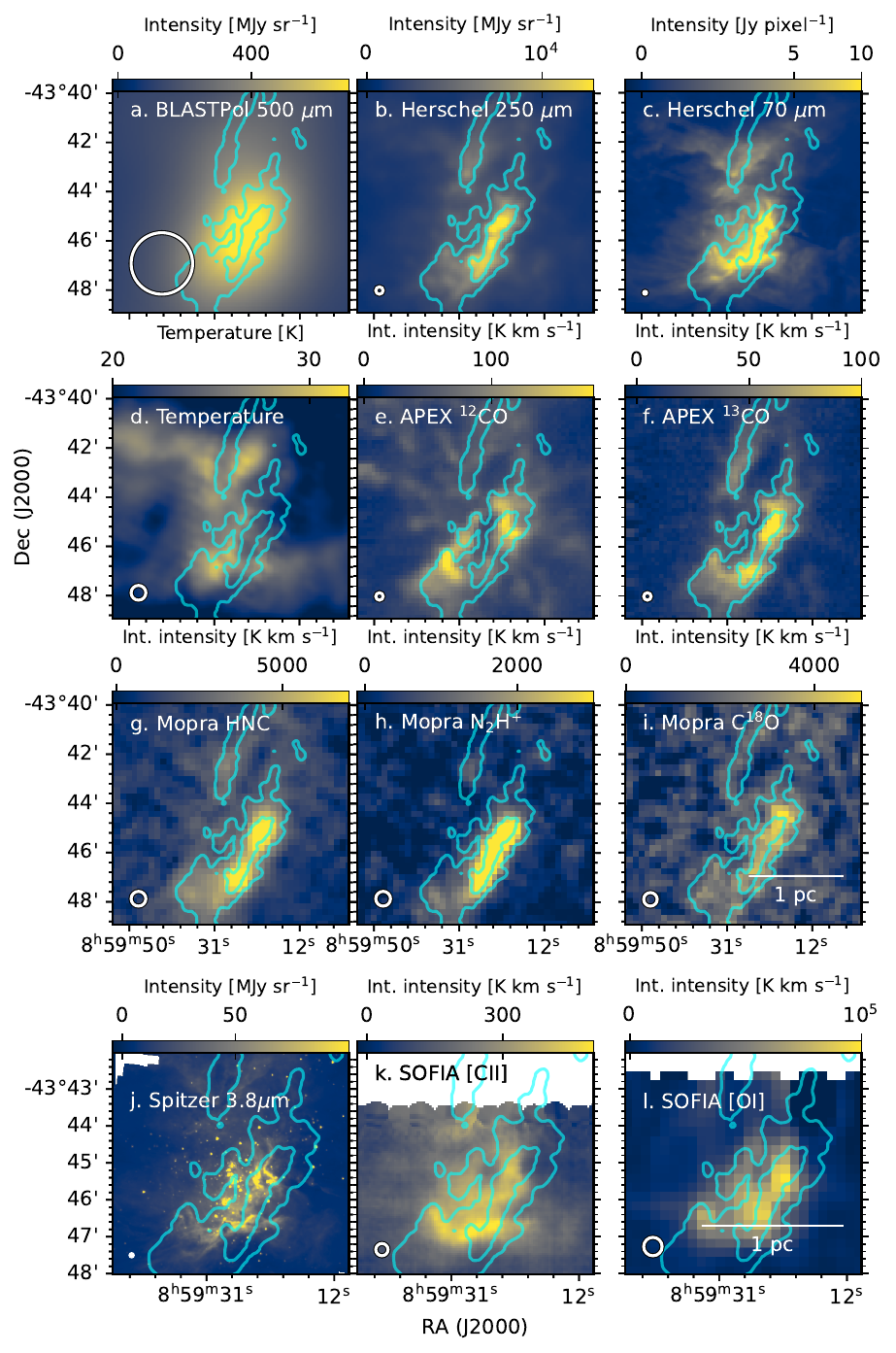}
    \caption{Maps of RCW 36 derived from different infrared and spectral line tracers. Panels $a$\textendash $i$ share the same RA and Dec axes, for which a scale bar is given in the bottom right of panel $i$. Panels $j$\textendash $l$ share another set of RA and Dec axes, for which a scale bar is given in panel $l$. The FWHM beam size is shown on the bottom left (beam and pixel sizes are given in Table \ref{tab:Obs}). For BLASTPol, the beam size is 2.5\arcmin{} and pixel size is 4.6\arcsec{}. The contours show the same \emph{Herschel}-derived column densities as Fig. \ref{fig:NewData}. The \emph{Herschel} far-IR flux maps in panels $b$ and $c$ are plotted with a log scale color map, the rest of the color maps are in linear scale. Panels $e$\textendash$i$ and $k$\textendash$l$ are integrated over the velocity ranges specified in Column 5 of Table \ref{tab:MolecularLines}.} 
    \label{fig:AuxData}  
\end{figure*}

In this study we use several types of data for the HRO analysis, which can be separated into three categories: polarized emission from dust (Section \ref{sec:Data_Pol}), thermal continuum emission from dust (Section \ref{sec:Data_DustEm}), and  spectroscopic lines (Section \ref{sec:Data_Spectra}), each of which are summarized in Table \ref{tab:Obs}. The observations from SOFIA HAWC+ and ALMA are presented for the first time. Figure \ref{fig:NewData} shows the new HAWC+ and ALMA data and Figure \ref{fig:AuxData} shows archival data. A detailed overview of the each dataset is provided in the following subsections.

\subsection{Dust Polarization Maps}
\label{sec:Data_Pol}

The linearly polarized intensity $P$ can be found from the linear polarization Stokes parameters $Q$ and $U$ using:

\begin{equation}
    \label{Pol_intensity}
    P = \sqrt{Q^2 + U^2},
\end{equation}

while the polarization fraction is given by $p = P/I$, where $I$ is the total intensity. 
The polarized intensity and polarization fraction are both constrained to be positive quantities which results in a positive bias at low total intensities \citep{1945BSTJ...24...46R, 1962AdA&A...1..289S}. This can be corrected for with `debiasing' \citep{1974ApJ...194..249W}, using:
\begin{equation}
    \label{Eqn:debias}
    P_{db} = \sqrt{Q^2 + U^2 - \frac{1}{2}(\delta Q^{2} + \delta U^2)},
\end{equation}

where $\delta Q$ and $\delta U$ are the measurement uncertainties in $Q$ and $U$, respectively. The polarization angle $\hat{E}$ can be calculated using $\hat{E}=0.5\,\mathrm{arctan}(U,Q)$. The orientation of the plane-of-sky magnetic field \Bfield\ is then inferred to be orthogonal to  $\hat{E}$ such that:

\begin{equation}
    \label{Eqn:Bfield_QU}
    \hat{B}_{\perp}=\frac{1}{2}\mathrm{arctan}(U,Q) + \frac{\pi}{2},
\end{equation}

where an angle of 0$^{\circ}$ points towards Equatorial north and increases eastward. 

\subsubsection{SOFIA HAWC+}
\label{sec:SOFIA_reduction}
In this paper we publish, for the first time, observations of RCW 36 using publicly available archival data from HAWC+ \citep{2018JAI.....740008H}, the far-IR polarimeter onboard SOFIA. The RCW 36 region was observed by SOFIA/HAWC+ on 6 and 14 June 2018 as part of the Guaranteed Time Observing (GTO) program (AOR: 70\_0609\_12), in both Band C (89 $\mu$m) and Band E (214 $\mu$m), at nominal angular resolutions at FWHM of 7.8\arcsec{} and 18.2\arcsec{}, respectively. The observations were done using the matched-chop-nod method \citep[described in][]{2000PASP..112.1215H} with a chopping of frequency of 10.2 Hz, chop angle of 112.4$^{\circ}$, nod angle of -67.5$^{\circ}$ and chop throw of 240\arcsec{}. Each observing block consisted of four dithered positions, displaced by 12\arcsec{} for Band C and 27\arcsec{} for Band E. The total observation time for Band C and Band E was 2845s and 947s, respectively. The total intensity maps for each band are shown in the top row of Figure \ref{fig:NewData}.

To reduce this data, we used the HAWC+ Data Reduction Pipeline which is described in detail in \cite{2019ApJ...882..113S} and \cite{2021ApJ...918...39L} and summarized here as follows. The pipeline begins by demodulating the data and discarding any data points affected by erroneous telescope movements or other data acquisition errors. This demodulated data is then flat-fielded to calibrate for gain fluctuations between pixels and combined into four sky images per independent pointing. The final Stokes $I$, $Q$ and $U$ maps are generated from these four maps after performing flux calibrations accounting for the atmospheric opacity and pointing offsets. Next, the polarization is debiased (see Equation \ref{Eqn:debias}) and the polarization percentage and the polarization angle are calculated.  
A $\chi^{2}$ statistic is then computed by comparing the consistency between repeated measurements to estimate additional sources of uncertainties such as noise correlated across pixels which can lead to an underestimation of errors in the $IQU$ maps \citep{2011ApJ...732...97D, 2013ApJ...770..151C}. This underestimation can be corrected for using an Excess Noise Factor (ENF) given by $\mathrm{ENF} = \sqrt{\chi^{2}/\chi_{\mathrm{theo}}}$ where $\chi_{\mathrm{theo}}$ is theoretically expected and $\chi^{2}$ is measured. The ENF is estimated in the HAWC+ Reduction Pipeline by fitting two parameters $I_{0}$ and $C_{0}$ using

\begin{equation}
    \label{Eqn:ENF}
    \mathrm{ENF} = C_{0}\sqrt{1 + \left(\frac{I}{5I_{0}}\right)^{2}},
\end{equation}

where $I$ is the total intensity (Stokes $I$) of a pixel in units of Jy pixel$^{-1}$. The errors of the $IQU$ maps are then multiplied by the ENF. The values of $I_{0}$, $C_{0}$ used for $I$, $Q$, and $U$ maps in Band C and E are summarized in Table \ref{tab:HAWC_ENF}. We note that the Stokes $I$ errors for Band E were a special case where the Pipeline ENF fitting routine failed (giving a value of $I_{0}=0$ which resulted in a non-physical ENF), likely due to large intensities from bright emission. In this case we forced the ENF to be about 1 (by manually setting $I_{0}$ to be a sufficiently large number such as 100 and $C_{0}=1$) such that the errors were neither under nor overestimated. 

After correcting the errors, the pipeline rejects any measurements falling below the 3-$\sigma$ cutoff in the degree of polarization $p$ to associated uncertainty $\sigma_{p}$ ($p < 3\sigma_{p}$) which roughly  corresponds to a 10$^{\circ}$ uncertainty in the polarization angle. 

After running the pipeline, we also applied a 3-$\sigma$ signal-to-noise threshold on the total intensity flux and polarized flux to further remove noisy polarization vectors. As a final diagnostic we checked for potential contamination from the reference beam position due to dithering the data in Chop-Nod polarization observations \citep[following the method described in the Appendix section of][]{1997ApJ...487..320N}. For this test we used \emph{Herschel} far-IR intensity maps (described in Section \ref{Sec:HerschelData}) since they cover both the RCW 36 region and the surrounding Vela C cloud-scale region, which include the HAWC+ chop reference beam positions that are outside the HAWC+ maps. We used \emph{Herschel} maps of comparable wavelengths to the HAWC+ data (PACS 70 $\mu$m to compare with HAWC+ 89 $\mu$m and PACS 160 and SPIRE 250 $\mu$m to compare with HAWC+ 214 $\mu$m) and found the ratio of the total intensity of the HAWC+ region compared to the chop reference beam region in the \emph{Herschel} data. To estimate the ratio of polarization flux, we conservatively assume that there is a 10\% polarization at the reference beam positions and remove points where the estimated polarized flux in the reference beam is more than 1/3 times the polarized flux in the HAWC+ map.

\begin{table}
    \caption{Excess Noise Factor (ENF) fitting parameters $I_{0}$ and $C_{0}$ found by the HAWC+ Data Reduction Pipeline for Stokes $I$, $Q$, $U$ Maps in Band C and E (see Eq. \ref{Eqn:ENF}).}
    \label{tab:HAWC_ENF}
    \centering
    \begin{tabular}{cccc}
        \hline
        Band & Stokes & $I_{0}$ & $C_{0}$ \\ 
        \hline \hline
        \multirow{3}{4em}{\centering{C} \\ \centering{89} $\mu$m} & $I$ & 0.332 & 25.929 \\
        & $Q$ & 0.716 & 1.723 \\
        & $U$ & 0.661 & 1.769 \\
        \hline
        \multirow{3}{4em}{\centering{E} \\ \centering{214 $\mu$m}} & $I$
        \footnote{Fitting routine failed and values were manually set to get and ENF $\sim 1$ (see text for details).} 
        & 100 & 1 \\
        & $Q$ & 1.067 & 1.544 \\
        & $U$ & 0.696 & 1.739 \\
        \hline
    \end{tabular}
\end{table}

\subsection{Dust Emission Maps}
\label{sec:Data_DustEm}
 
\subsubsection{ALMA}
In this work, we present two new interferometric data sets from the Atacama Large Millimeter/submillimeter Array (ALMA) and Atacama Compact Array (ACA). 

The first dataset includes observations of dust continuum and line emission in Band 6 (1.1\textendash1.4 mm) using the ACA with 7-m dishes (ID 2018.1.01003.S, Cycle 6, PI: Fissel, Laura). The observations took place from 22 April to 14 July of 2019 with a continuum sensitivity of 2.15 mJy beam$^{-1}$. The configuration resulted in a minimum baseline of 9 m and a maximum baseline of 48 m. The angular resolution is approximately 4.9\arcsec{} and the maximum recoverable scale is 28\arcsec{}, which corresponds to spatial scales ranging from $\sim$0.02 to 0.12 pc (4410 to 25200 au) using a distance estimate of $\sim 900$ pc for Vela C from \cite{2020A&A...633A..51Z}. The imaged area was 108\arcsec{} $\times$ 324\arcsec{}, with 87 mosaic pointings and a mosaic spacing of 21.8\arcsec{}. The average integration time was 20 minutes per mosaic pointing.  

The second ALMA program used the 12-m array in the C43-1 configuration to observe both polarization mosaics of some of the dense cores identified in the ACA data, as well as larger spectroscopic and continuum observations with the same correlator configuration as the ACA observations (ID 2021.1.01365.S, Cycle 8, PI: Bij, Akanksha). The spectroscopic and continuum observations took place on 19\textendash24 March 2022 with a continuum sensitivity of 0.2 mJy beam$^{-1}$. The configuration resulted in a minimum baseline of 14.6 m and a maximum baseline of 284 m. The angular resolution is approximately 1.2\arcsec{} and the maximum recoverable scale is 11.2\arcsec{}, which corresponds to $\sim$0.0052 to 0.05 pc (1080 to 10080 au) resolution at the distance to Vela C. The imaged area was 42\arcsec{} $\times$ 85\arcsec{} in size, with 26 mosaic pointings and a mosaic spacing of 12.3\arcsec{}. The average integration time was 9.5 minutes per mosaic pointing. 

In this work, we present only the total intensity dust continuum maps from both datasets, as delivered by the QA2 (Quality Assurance 2) process, with no further data reduction performed. These maps are shown in the bottom row of Fig. \ref{fig:NewData}. We do not analyze the spectral data and polarization mosaics from these observations in this work, as further data reduction is required and left for future studies. 

\subsubsection{Herschel SPIRE and PACS}
\label{Sec:HerschelData}
To study the cloud structures probed by thermal dust emission, we use publicly available archival maps from the \emph{Herschel} Space Observatory, which observed Vela C on 18 May 2010 \citep{2011A&A...533A..94H} as part of the \emph{Herschel} OB Young Stars (HOBYS) key programme \citep{2010A&A...518L..77M}. The observations were conducted using the SPIRE instrument at 500, 350, and 250 $\mu$m \citep[with FWHM angular resolutions of 36\arcsec{}, 25\arcsec{}, and 18\arcsec{}, respectively;][]{2010A&A...518L...3G}, and the PACS instrument at 70 and 160 $\mu$m \citep[with resolutions of 8\arcsec{} and 13\arcsec{}, respectively;][]{2010A&A...518L...2P, 2012A&A...539A.156G}. Additionally, we include a \emph{Herschel}-derived temperature map at an angular resolution of 36\arcsec{} and a 18\arcsec{} column density map. The column density map is derived from a spectral energy distribution fit to the 160, 250, 350 and 500 $\mu$m flux maps, following the procedure described in detail in Appendix A of \citealt{2013A&A...550A..38P}. We show the 250 $\mu$m, 70 $\mu$m, and temperature maps in panels b, c, and d of Figure \ref{fig:AuxData}, respectively. 

\subsubsection{Spitzer IRAC}
To trace warmer dust grains, we use archival mid-IR maps from the \emph{Spitzer} Space Telescope, obtained from the publicly available ISRA NASA/IPAC Infrared Science Archive\footnote{Available at https://irsa.ipac.caltech.edu/about.html}. \emph{Spitzer} observed RCW 36 in May 2006, employing the four-channel camera IRAC to capture simultaneous broadband images at channels 1\textendash4, covering bands centered at 3.6, 4.5, 5.8, and 8.0 $\mu$m, respectively \citep{2004ApJS..154...10F}. IRAC uses two 5.2\arcmin{} $\times$ 5.2\arcmin{} fields of view, where one field simultaneously images at 3.6 and 5.8 $\mu$m and the other at 4.5 and 8.0 $\mu$m. All four detector arrays are 256 $\times$ 256 pixels, with 1.2\arcsec{} square pixels. This dataset was published in \citep{2013A&A...558A.102E}. In this work, we only use data from 3.6 $\mu$m (channel 1) and 4.5 $\mu$m (channel 2) for our HRO analysis, as channels 3 and 4 have artifacts and saturation. Channels 1 and 2 have resolutions of 1.66\arcsec{} and 1.78\arcsec{}, respectively. We show the \emph{Spitzer} 3.6 $\mu$m map in panel j of Figure \ref{fig:AuxData}. 

\subsection{Atomic and Molecular Line Maps}
\label{sec:Data_Spectra}

\begin{table*}
\caption{Summary of Spectroscopic Data Cubes}
\label{tab:MolecularLines}
\centering
    \begin{threeparttable}
    \begin{tabular}{cccccc}
        \hline
        Instrument & Data Type & Rest Frequency & 
        $\Delta v\,\,^{a}$ &
        Single HRO$\,\,^{b}$ & Vel. HRO$\,\,^{c}$  \\
        & & (GHz) & (km s$^{-1}$) & $v_{0}$\textendash$v_{1}$ (km s$^{-1}$) & $v_{0}$\textendash$v_{1}$ (km s$^{-1}$) \\
        \hline \hline
        \multirow{2}{*}{APEX/LAsMA} & 
        $^{12}$CO (3--2) & 345.7960 & 0.2  & -20\textendash20 & 0\textendash10  \\
        & $^{13}$CO (3--2) & 330.5880 & 0.2  & -20\textendash20 & 0\textendash10 \\
        \hline      
        \multirow{2}{*}{SOFIA/upGREAT} &
         [C$\mathrm{_{II}}$]$^{3}$P$_{3/2} \rightarrow ^{3}$P$_{1/2}$ & 1897.4206 & 0.2 & -20\textendash20 & -5\textendash10 \\
         & [O$\mathrm{_{I}}$] $^{3}$P$_{1} \rightarrow ^{3}$P$_{2}$ & 4758.6104 & 0.2 & -20\textendash20 & 0\textendash10  \\
        \hline
        \multirow{2}{*}{Mopra} &
         HNC (1--0) & 90.6636 & 0.22 &  0\textendash10 & 0\textendash10   \\
         & N$_{2}$H$^{+}$ (1--0) & 93.1730 & 0.21 & -6\textendash10 & -6\textendash10  \\
         & C$^{18}$O (1--0) & 109.7822 &  0.18 & 0\textendash10 & 0\textendash10 \\
         \hline
    \end{tabular}
    \begin{tablenotes}
        \small
        \item $^{a}$ Channel velocity resolution for each molecular line cube.
        \item $^{b}$ The range over which the integrated intensity map is calculated using Eq. \ref{Eqn:TotalIntIntensity} in the Single Map HRO.
        \item $^{c}$ The bounds for the velocity slabs described in Eq. \ref{Eqn:VelocitySlabInt} used for the Velocity Dependent HRO.
        \item The systemic velocity for the Centre-Ridge is $\sim$7 km/s \citep{2013A&A...550A..50M}.
    \end{tablenotes}
    \end{threeparttable}
\end{table*}

The gas structure of the region is also of significant interest in understanding the dynamic importance of the magnetic field and how it has been affected by  stellar feedback. To this end, we use a myriad of archival spectroscopic line data to probe different chemical, thermal and density conditions within the RCW 36 region. Table \ref{tab:MolecularLines} summarizes the lines of interest, including their transitions, rest frequencies, velocity resolution, and the channels used to make the integrated intensity maps. For our analysis of the spectroscopic data, we use both an integrated intensity map for a wide velocity range (Column 5 of Table \ref{tab:MolecularLines}) as well as channel maps over narrower velocity ranges (Column 6 of Table \ref{tab:MolecularLines}). Our spectral data cube analysis is described in Sections \ref{sec:Methods_single} and \ref{sec:Methods_velocity}.

\subsubsection{SOFIA upGREAT FEEDBACK}
To trace the PDR, we use a 158 $\mu$m map of the \CII\ $^{3}$P$_{3/2} \rightarrow ^{3}$P$_{1/2}$ transition (at native resolution of 14.1\arcsec{}) and a 63 $\mu$m map of the \OI\ $^{3}$P$_{1} \rightarrow ^{3}$P$_{2}$ transition (at native resolution of 6.3\arcsec{}) from the SOFIA FEEDBACK C$^{+}$ legacy survey \citep{2020PASP..132j4301S}. The survey was conducted by the upGREAT (upgraded German REceiver for Astronomy at Terahertz frequencies) heterodyne spectrometer \citep{2018JAI.....740014R} onboard the SOFIA aircraft \citep{2012ApJ...749L..17Y} on 6 June 2019 from New Zealand. The upGREAT receivers use a low frequency array (LFA) to cover the 1.9-2.5 THz band with 14 pixels and a high frequency array (HFA) covering the 4.7 THz line with 7 pixels. The on-the-fly observing strategy and data reduction for the survey is discussed in \cite{2020PASP..132j4301S}. In this work, we use the data reduced by \cite{2022ApJ...935..171B} who smoothed the \CII\ map to 20\arcsec{} and \OI\ map to 30\arcsec{} to reduce noise. They also applied principal component analysis (PCA) to identify and remove systematic components of baseline variation in the spectra. Both maps are 14.4\arcmin{} $\times$ 7.2\arcmin{} in size, with a spectral binning of 0.2 km/s, for which the typical rms noise is 0.8$-$1.0 K for \CII\ and $\sim$0.8$-$1.5 K for \OI\ \citep{2022ApJ...935..171B}. The integrated intensity maps for \CII\ and \OI\ and have been integrated from -20 to +20 km/s are shown in panels k and l of Fig. \ref{fig:AuxData}, respectively.

\subsubsection{APEX LAsMA}
To trace the molecular gas regions in RCW 36, we use observations of $^{12}$CO (3--2) and $^{13}$CO (3--2) obtained on 27 September 2019 and 21 July 2021 with the heterodyne spectrometer LAsMA (Large APEX Sub-Millimetre Array), which is a 7 pixel receiver on the APEX telescope \citep{2006A&A...454L..13G}. The maps were scanned in total power on-the-fly mode and are sized 20\arcmin{} $\times$ 15\arcmin{}, with a beamsize of 18.2\arcsec{} at 345.8 GHz. \cite{2022ApJ...935..171B} reduced the data to produce the baseline-subtracted spectra presented here, which have a spectral resolution of 0.2 km/s, pixel size of 9.1\arcsec{} and rms noise of 0.45 K. The integrated intensity maps have been integrated from -20 to +20 km/s for $^{12}$CO and $^{13}$CO are shown in panels e and f of Fig. \ref{fig:AuxData}, respectively.

\subsubsection{Mopra}
In our analysis, we also utilize complementary molecular line surveys from the 22-m Mopra Telescope, which observed the large-scale dense gas over the entire Vela C molecular cloud from 2009 to 2013. In this work, we use only the (1--0) transitions of C$^{18}$O as well as HNC and N$_{2}$H$^{+}$ which were originally presented in \cite{2019ApJ...878..110F}. The C$^{18}$O observations were performed by scanning long rectangular strips of 6\arcmin{} height in the galactic latitude and longitude directions using Mopra's fast-scanning mode. The HNC and N$_{2}$H$^{+}$ observations used overlapping 5\arcmin{} square raster maps. The data reduction procedure, performed by \cite{2019ApJ...878..110F}, includes bandpass correction using off-source spectra, bandpass polynomial fitting and Hanning smoothing in velocity. The resulting FWHM angular resolution and velocity resolution is 33\arcsec{} and 0.18 km/s for C$^{18}$O, and 36\arcsec{} and $\sim$0.2 km/s for both HNC and N$_{2}$H$^{+}$. The integrated intensity maps for HNC, N$_{2}$H$^{+}$ and C$^{18}$O have been integrated over the velocity range given in Column 5 of Table \ref{tab:MolecularLines} and are shown in panels g--i of Fig. \ref{fig:AuxData}, respectively.

\section{RCW 36 Structure}
\label{sec:Structure}
    \begin{figure*}
      \centering
      \includegraphics[width=\hsize]{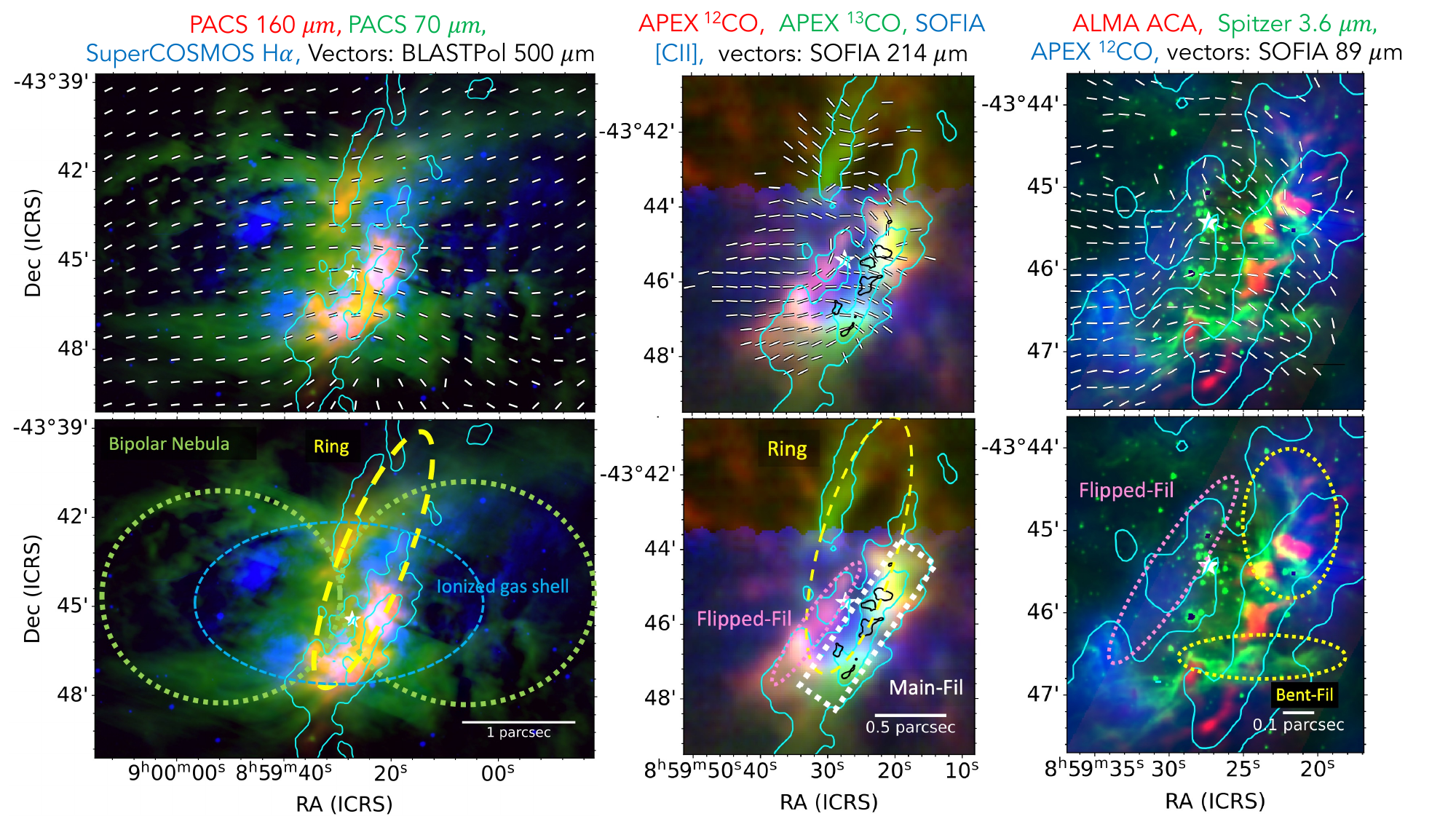}
      \caption{Three color-composite images of the RCW 36 region on different scales. All images use either linear intensity or log intensity to highlight emission structures. The white star marker in all panels shows the location of the brightest
      O9 V star in RCW 36 \citep[Object ID number 1 from][and number 462 from \citealt{2005A&A...440..121B, 2006A&A...455..561B}]{2013A&A...550A..50M}. The cyan contours represent the same \emph{Herschel}-derived column densities as Figure \ref{fig:NewData}. \textit{Left: Cloud-scales}. The red-blue-green colours are the same for both top and bottom panels with \emph{Herschel}/PACS 160 $\mu$m (red), \emph{Herschel}/PACS 70 $\mu$m (green), and SuperCOSMOS \Halpha\ (blue). The top panel vectors show the magnetic field orientation inferred from BLASTPol 500 $\mu$m polarized data. The bottom panel has labels for the bipolar nebula (green dotted), ring (yellow dashed) and ionized gas shell (blue dashed). \textit{Middle: Filament-scales} The red-blue-green colours are the same for both top and bottom panels with APEX $^{12}$CO integrated intensity(red), APEX $^{13}$CO integrated intensity (green), and SOFIA \CII\ integrated intensity (blue). Note that the \CII\ emission does not cover the northern part of the image. The top panel vectors show the magnetic field orientation inferred from HAWC+ 214 $\mu$m (Band E)  polarized data. The bottom panel labels the Main-Fil (white dashed) and Flipped-Fil (pink dotted). The Main-Fil contains five dense star-forming clumps outlined by black contours showing ALMA ACA continuum at an intensity of 0.018 Jy beam$^{-1}$. The Flipped-Fil corresponds to the region where the magnetic field orientation appears to abruptly change by almost 90$^{\circ}$. \textit{Right: Clump-scales} The red-blue-green colours are the same for  both top and bottom panels with ALMA ACA 1.1-1.4 mm continuum (red), \emph{Spitzer} 3.6 $\mu$m intensity (green), and APEX $^{13}$CO integrated intensity (blue). Top panel vectors show the magnetic field orientation inferred from HAWC+ 89 $\mu$m (Band C) polarized data. The Bottom panel shows the North and South Bent-Fils (yellow-dotted). Each of the lower panels also shows a scale bar.
      }
      \label{figure:RGBScale}
    \end{figure*}

In this section, we give an overview of the morphological structure and magnetic field geometry of RCW 36 on varying spatial scales based on previous studies of the region, as well as inferences based on our observations. Figure \ref{figure:RGBScale} showcases various continuum and polarization observations for RCW 36 that will be used to describe its general structure in the following subsections.  

\subsection{Cloud-Scale}
On cloud scales of $> 1$ pc, the RCW 36 region is located within the Vela C giant molecular cloud. Vela C consists of a network of filaments, ridges and nests, which were identified by \cite{2011A&A...533A..94H} using \emph{Herschel} data. The densest and most prominent of the ridges is the Centre-Ridge, with column densities of \Av\ $> 30$, and a length of roughly $\sim$10 pc \citep{2011A&A...533A..94H}. The Center-Ridge contains the RCW 36 region. It has a bipolar nebula morphology \citep{2013A&A...550A..50M} with two fairly symmetric lobes oriented in the east-west direction that are traced well by the green PACS 70 $\mu$m emission in the lower-left panel of Figure \ref{figure:RGBScale} (see also the dotted green ellipses). This bipolar nebula is roughly centered around a young (1.1 $\pm$ 0.6 My) massive cluster with two late-type O-type stars and $\sim$350 members \citep{2013A&A...558A.102E}. The position of the most massive star (spectral type O9V) is indicated by a white star-shaped marker in Fig. \ref{figure:RGBScale}. The ionizing radiation from this cluster is powering an expanding \HII\ gas shell traced by H$\alpha$ (shown in blue, Fig. \ref{figure:RGBScale}, left panel) \citep{2013A&A...550A..50M}. 

Bipolar \HII\ regions are of great interest because, though they have been observed in other high-mass star forming regions such as S106 \citep{2018A&A...617A..45S} and G319.88+00.79 \citep{2015A&A...582A...1D}, they seem to be more rare than single \HII\ bubbles \citep[e.g.,][]{2006ApJ...649..759C, 2010A&A...523A...6D, 2011ApJS..194...32A, 2012ApJ...755...71K, 2018A&A...617A..67S}.

Within the bipolar cavities, \cite{2022ApJ...935..171B} identified blue-shifted \CII\ shells with a velocity of 5.2$\pm$0.5 km/s, likely driven by stellar winds from the massive cluster. Additionally, they find diffuse X-ray emission (observed with the Chandra X-ray Observatory) in and around the RCW 36 region which is tracing hot plasma created by the winds. However, \cite{2022ApJ...935..171B} estimate that the energy of the hot plasma is 50--97\% lower than the energy injected by stellar winds and reason that the missing energy may be due to plasma leakage, as has been previously suggested for RCW 49 \citep{2021AAS...23713703T}. 

The magnetic field geometry on $> 1$ pc cloud scales is traced by the BLASTPol 500 $\mu$m polarization map \citep{2016ApJ...824..134F} which has a FWHM 2.5\arcmin{} resolution, corresponding to 0.65 pc at the distance of Vela C. The BLASTPol magnetic field orientation is shown by vectors in top-left panel of Fig. \ref{figure:RGBScale}, which follow a fairly uniform east-west morphology that is mostly perpendicular to the orientation of the dense ridge. However, around the north and south `bends' of the bipolar structure, the magnetic field lines also appear to bend inward towards the center, following the bipolar shape of the structure. 

\subsection{Filament-Scale}
The middle panels of Figure \ref{figure:RGBScale}  highlight structures on filament scales of $\sim0.1\textendash1$ pc and the cyan contours in all panels represent \emph{Herschel}-derived column densities to show the filament. At the waist of the bipolar nebula, \cite{2013A&A...550A..50M} identify a ring-like structure that extends $\sim1$ pc in radius and is oriented perpendicular to the bipolar nebula lobes, in the north-south direction (labeled in yellow in both the left and middle panels of Fig. \ref{figure:RGBScale}). The majority of the dense material, as traced by the column density contours, is contained within this ring. \cite{2013A&A...550A..50M} also model the kinematics of the ring and find that the north-eastern (NE) half is mainly blue-shifted while the south-western (SW) half is red-shifted, consistent with an expanding cloud with speeds of 1\textendash2 km/s.
To trace the ionized gas, i.e. \HII\, we use archival \Halpha\ data from the SuperCOSMOS H-alpha Survey \citep[SHS;][]{2001MNRAS.326.1279H, 2005MNRAS.362..689P}. From the SuperCOSMOS map, we note that the eastern side of the ring is seen in \Halpha\ absorption, signifying that it is in front of the ionizing gas and associated massive star cluster. Whereas, \Halpha\ emission is seen across the western region of the ring and therefore this part of the ring is likely behind the cluster. 

The highest column density contours are observed within the SW half of the ring, where most of the next-generation star formation appears to be taking place. We henceforth refer to this region as the `Main-Fil' (labeled in white, middle panel, Fig \ref{figure:RGBScale}). \cite{2012A&A...548L...6H} estimate that the mass per unit length of the Main-Fil region is 400$\pm$85 M$_{\odot}$ pc$^{-1}$. The Main-Fil is seen to host multiple star-forming cores and/or clumps which are shown by the black ALMA Band 6 continuum contours in the middle panel of Fig \ref{figure:RGBScale}.

Several diffuse filamentary structures traced by $^{12}$CO and \CII\ (shown in red and blue, respectively in middle panel of Fig. \ref{figure:RGBScale}) can also be observed in the ambient cloud surrounding the ring. \cite{2022ApJ...935..171B} have reasoned that due to the curved shape of these filaments, they are not part of the larger expanding \CII\ shells but may have been low density filaments originally converging towards the centre dense ridge (similar to the converging filaments seen in Musca, B211/3, and DR 21; \citealt{2008ApJ...680..428G, 2010A&A...520A..49S, 2013A&A...550A..38P, 2016A&A...590A.110C, 2023ApJ...951...39B}) that have instead been swept away at velocities $>$ 3 km/s due to stellar feedback.

The magnetic field geometry on filament scales, traced by SOFIA/HAWC+ 214 $\mu$m is fairly consistent with the E-W geometry seen on cloud scales, with some interesting exceptions. The most striking deviation is the region located just east of the ionising stars, hereafter referred to as `Flipped-Fil' (labeled in Fig \ref{figure:RGBScale}, middle and right panels). Here the magnetic field morphology, as traced by SOFIA/HAWC+, deviates from the E-W trend and abruptly flips almost by 90$^{\circ}$ to follow a more N-S configuration. This geometry appears to follow the elongation of a lower density filament traced by $^{12}$CO (red, middle panel), \CII\ (blue, middle panel), and $^{13}$CO emission  (blue, right panel).

\subsection{Core-Scale}
\label{sec:Clump-Scale}
The right panels of Figure \ref{figure:RGBScale} show emission on sub-clump and core scales ($<$ 0.1 pc) in RCW 36. These data reveal complex substructure within the Main-Fil region. The Main-Fil is clumpy with several bright rims, voids, and pillar-like structures identified by \cite{2013A&A...550A..50M} (see their Figure 3). This matches our ALMA band 6 continuum observations as shown by the five near-round clumps and associated elongated pillars, as seen in the right panel Fig. \ref{figure:RGBScale}. 

\cite{2013A&A...550A..50M} recognized that the bright rims appear near the end of the pillar-like structures. The bright-rims are traced in the right panel of Fig.\ref{figure:RGBScale} by \emph{Spitzer}/IRAC 3.6 $\mu$m emission, which mainly traces hot dust, found at the edges of the PDR \citep{2013A&A...550A..50M}. These rims appear to wrap around the cold ALMA clumps, without covering them completely, in a manner resembling bow-shocks (though actual bow-shocks are unlikely in this region). These bright rims are of great interest in this work and are therefore collectively labeled as `Bent-Fils' as they will be referred to in later sections. There is a prominent northern Bent-Fil and southern Bent-Fil shown by the two yellow-dotted ovals in the lower right panel of Fig. \ref{figure:RGBScale}. The curved morphology of the Bent-Fils is noted by \cite{2013A&A...550A..50M} to be likely due to tracing the inner border of the dense ring which is being progressively photoionzed by the star cluster. Interestingly, the HAWC+ magnetic field morphology seems to follow the Bent-Fil features, deviating once again from the general E-W cloud scale magnetic field. 

\section{Methods}
\label{sec:Methods}

\subsection{Histogram of Relative Orientations}

In this section we discuss the procedure of the Histogram of Relative Orientation (HRO) method \citep[see][for a more detailed description]{2013ApJ...774..128S, 2017A&A...607A...2S} which computes the relative angle 
$\phi(x,y)$ between the gradient vector field of a structure map $M(x,y)$ and the plane-of-sky magnetic field orientation \Bfield\ at each pixel. The steps for this procedure are outlined in the following subsections.

\subsection{Preparing the Structure Map}

In this work, we apply two different methods to obtain a two-dimensional structure map $M(x,y)$. In the first approach, henceforth referred to as Single Map HRO (described further in \ref{sec:Methods_single}), we compare the orientation of local structure at every location using one map $M(x,y)$, for each of the datasets listed in Table \ref{tab:Obs}, to the magnetic field orientation measured by dust polarization as was done by previous HRO studies \citep[e.g.]{2016A&A...586A.138P, 2017A&A...603A..64S, 2019ApJ...878..110F, 2021ApJ...918...39L}. In the second approach, applied only to the spectroscopic data cubes listed in Table \ref{tab:MolecularLines}, we slice the spectral line cube into multiple velocity slabs $v_{i}$ and compare the orientation of structures in the integrated intensity of each slab $M(x,y)_{i}$ to the inferred magnetic field. This quantifies the relative alignment as a function of line-of-sight velocity, and will therefore be referred to as the Velocity Dependent HRO (see \ref{sec:Methods_velocity}).

\subsubsection{Single Map HRO}
\label{sec:Methods_single}
The dust emission, column density, and temperature maps are already in the 2-D spatial format and are thus used directly as the $M(x,y)$ structure map in the Single Map HRO analysis. For the spectroscopic cubes, we generate a single integrated line intensity map as $M(x,y)$ (following the procedure of \citealt{2019ApJ...878..110F}). To calculate the integrated line intensity, a velocity range $v_{0}$--$v_{1}$ is selected for each molecular line over which the the radiation temperature $T_{R}$ in a given velocity channel $v$ is integrated, using

\begin{equation}
    \label{Eqn:TotalIntIntensity}
    M(x,y) = \int_{v_{0}}^{v_{1}} T_{R} \, dv.
\end{equation}

The velocity ranges used to calculate the integrated intensity map in the Single Map HRO analysis for each spectral cube are specified in Column 5 of Table \ref{tab:MolecularLines}.

\subsubsection{Velocity Dependent HRO}
\label{sec:Methods_velocity}

Additionally for the molecular line data, we also perform a Velocity Dependent HRO analysis. Here we slice a selected velocity range $v_{0}$--$v_{1}$(specified in Column 6 of Table \ref{tab:MolecularLines}) into narrower velocity slabs with a width of 1 km/s. We increment the center velocities $v_{i}$ of each slab by 0.5 km/s, such that $v_{i}$=\{$v_{0}+0.5, \, v_{0}+1,  \dots, \, v_{1}-1, \, v_{1}-0.5$\}. For every velocity $v_{i}$ in the set, we generate an integrated intensity map M$(x,y)_{i}$ using:

\begin{equation}
    \label{Eqn:VelocitySlabInt}
    M(x,y)_{i} = \int_{v_{i}-0.5}^{v_{i}+0.5} T_{R} \, dv.
\end{equation}

The 1 km/s width of the slabs are chosen to be roughly a factor of 5 larger than the $\sim0.2$ km/s velocity resolution of the data cubes (listed in Column 4 of Table \ref{tab:MolecularLines}) such that enough velocity channels are included in the integrated intensity map. This ensures that there is sufficient signal-to-noise in each slab and small local fluctuations are averaged over. The HRO analysis is then repeated for each $M(x,y)_{i}$ in the set. 

\subsection{Projection and Masking}
\label{sec:ProjandMask}

To directly compare the structure map $M(x,y)$ and the plane-of-sky magnetic field map \Bfield\, the next step is to ensure that both maps share the same spatial coordinate grid such that there is one-to-one mapping between the pixels. To do this, the map with the coarser pixel scale is projected on to the grid of the map with the finer pixel scale (i.e., if the $M(x,y)$ map has a lower resolution than the \Bfield\ map then $M(x,y)$ is projected on the \Bfield\ map). The pixel sizes of each dataset are given in Table \ref{tab:Obs} for comparison. When the \Bfield\ data has lower resolution of the two, rather than directly projecting the orientation of the magnetic field lines inferred from Eq. \ref{Eqn:Bfield_QU}, we instead project the Stokes $Q$ and $U$ intensity maps separately and then recalculate the inferred \Bfield. This avoids an incorrect assignment of vector orientation angles to the newly sized pixels. 

Next, a 3-$\sigma$ signal-to-noise cut is applied to the data points in the structure map $M(x,y)$. For the Single Map HRO analysis, all of the $M(x,y)$ maps of the RCW 36 region are above this threshold for every point that overlaps with the \Bfield\ data, with the exception of the ALMA continuum maps for which the signal is concentrated near the dense clumps. For the Velocity Dependent HRO analysis, the $M(x,y)_{i}$ integrated intensity map of each velocity slab is masked individually. 

\subsection{Calculating the Relative Orientation Angle}

To determine the orientation of elongated structures in $M(x,y)$, we calculate the direction of the iso-contours $\psi (x,y)$ (which is by definition perpendicular to the gradient vector field $\nabla$M), given by: 

\begin{equation}
    \psi=\mathrm{arctan}\left( \frac{\delta M / \delta x}{\delta M / \delta y} \right),
\end{equation}

where $\psi$ is calculated at each pixel $(x,y)$. The partial derivatives are calculated by convolving $M(x,y)$ with Gaussian derivative kernels G, using 

\begin{equation}
   \label{Eqn:Gauss_smooth}
   \frac{\delta M}{\delta x} = M(x,y) \ast \frac{\delta}{\delta x} G(x,y),   
\end{equation}

and similarly $\delta\mathrm{M}/\delta y = \mathrm{M}(x,y) \star \delta G(x,y) / \delta y$. This reduces noise and avoids erroneous relative angle measurements due to map pixelization. The size of the Gaussian kernels in angular units $\theta_{G}$ is chosen to be one third of the FWHM angular resolution $\theta_{beam}$ of the $M(x,y)$ map, using $\theta_{G} = \theta_{beam} / 3$. If this kernel size $\theta_{G}$ is less than 3 pixels, then a minimum kernel size of 3 pixels is used instead. A summary of all the kernel sizes in angular units $\theta_{G}$ and pixel units $l_{G}$ is provided in Columns 5 and 6, respectively, of Table \ref{tab:Obs}. The same smoothing lengths listed in Table \ref{tab:Obs} for the molecular line data are applied for the Velocity Dependent HRO analysis.

The relative angle $\phi(x,y)$ between the iso-contour direction $\psi(x,y)$ and the plane-of-sky magnetic field \Bfield\ can then be computed with:

\begin{equation}
    \label{Eqn:RelativeAngle}
    \phi \equiv \mathrm{arctan}\left( \frac{|\psi \times \hat{B}_{\perp}|}{\psi \cdot \hat{B}_{\perp}} \right).
\end{equation}

The resulting $\phi$ falls within the range [0$^{\circ}$, 180$^{\circ}$], but since $\phi$ measures only the orientation and not direction, the angles $\phi$ and 180$-\phi$ are redundant. The range can therefore be wrapped on [0$^{\circ}$, 90$^{\circ}$] as we are only concerned with angular distance, such that $\phi=0^{\circ}$ (and equivalently $\phi=180^{\circ}$ before wrapping) corresponds to the local structures being aligned parallel relative to the magnetic field orientation, while $\phi=90^{\circ}$ corresponds to perpendicular relative alignment. A histogram can then be used to combine the relative angle measurements across all pixels in order to summarise the overall trend within the map. We place the $\phi(x,y)$ measurements into 20 bins over [0$^{\circ}$, 90$^{\circ}$], where each bin is of 4.5$^{\circ}$ in size.

\subsection{Projected Rayleigh Statistic}
\label{sec:PRS}

While the histogram is a useful tool for checking if there is a preference towards a particular relative angle, we can go a step further and quantify the statistical significance of such a preference by calculating the Projected Rayleigh statistic (PRS) (as described in \citealt{2018MNRAS.474.1018J}). The PRS is a modified version of a classic Rayleigh statistic which tests for a uniform distribution of angles using a random walk. The classic Rayleigh statistic characterizes the distance $Z$ from the origin if one were to take unit-sized steps in the direction determined by each angle. Given a set {$\theta_{i}$} of n angles within the range $[0^{\circ}, 360^{\circ}]$, this distance $Z$ can be calculated as follows:

\begin{equation}
    Z = \frac{(\Sigma_{i}^{n} cos\theta_{i})^{2} + (\Sigma_{i}^{n} sin\theta_{i})^{2}}{n},
\end{equation}

where $n$ is the number of data samples. To use the set of relative angles $\phi(x,y)$ in the range $[0^{\circ}, 90^{\circ}]$ determined from HROs, we can map each angle $\theta=2\phi$. The range of possible $Z$ then is $[0, n]$, where $Z=0$ is expected if the angles {$\theta_{i}$} are distributed randomly and $Z=n$ is expected if all angles are the same. Any significant deviation from the origin would signify that the angles $\theta_{i}$ have a directional preference and are non-uniform. While this statistic is useful for testing for uniformity, it cannot differentiate between the preference for parallel versus perpendicular alignment, which is what we would like to measure in the context of HROs. To achieve this, \cite{2018MNRAS.474.1018J} modify this statistic by calculating only the horizontal displacement \Zprime\ in the hypothetical random walk:

\begin{equation}
\label{Eqn:UnCorrPRS}
    Z'_{x} = \frac{\Sigma_{i}^{n} cos\theta_{i}}{\sqrt{n/2}}.
\end{equation}

Now a parallel relative angle $\phi_{i}=0$ will map to $\theta=2\phi=0$ and give a positive $cos(0)=1$ contribution to \Zprime, while a perpendicular relative angle $\phi_{i}=\pi/2$ will map to $\theta=2\phi=\pi$ and give a negative $cos(\pi)=-1$ contribution to \Zprime. Therefore,  a statistic of \Zprime\ $\gg$ 0 indicates strong parallel alignment, while \Zprime\ $\ll$ 0 indicates strong perpendicular alignment. Since the value of \Zprime\ is within the range $[-\sqrt{2n}, +\sqrt{2n}]$, the statistic can be normalized by $\sqrt{2n}$ to give a measure of the degree of alignment:

\begin{equation}
    \label{Eq:normalizedUncorrPRS}
    \Tilde{Z}'_{x} = \frac{Z'_{x}}{\sqrt{2n}},
\end{equation}

where a value of $\Tilde{Z}'_{x}=\pm1$ would correspond to perfectly parallel or perpendicular alignment. 

If the $n$ data points are independent, then \Zprime\ should have an uncertainty of 1. However, most of the maps used in this HRO study are over-sampled and adjacent pixels are not entirely independent. Since the magnitude of \Zprime\ is proportional to the number of data points $n^{1/2}$ and the relative alignment $\phi$ is measured at each pixel $(x,y)$, oversampling within the map can result in a misleadingly large \Zprime\ magnitude. To determine the statistical significance of \Zprime\, we follow the methodology in \cite{2019ApJ...878..110F} and correct for oversampling by repeating the HRO analysis on 1000 independent white noise maps $M_{WN}$, smoothed to the same resolution as $M(x,y)$, and compared to the magnetic field orientation. 

The white noise maps $M_{WN}$ are generated to be the same size as the real data ($M(x,y)$), using Gaussian noise with a mean and standard deviation of $M(x,y)$. In these maps the orientation of the gradient will be a uniformly random distribution. The white noise $M_{WN}$ map then follows the same projection procedure that was applied to $M(x,y)$ in Section \ref{sec:ProjandMask} followed by the same mask which was applied to the real data $M(x,y)$. We calculate the corresponding PRS, $Z'_{WN}$ for each white noise map and determine the mean $\langle Z'_{WN} \rangle$ and standard deviation $\sigma_{Z'_{WN}}$ of the PRS in the 1000 runs. The value of $\sigma_{Z'_{WN}}$ estimates the amount of oversampling in the map. We can then correct the PRS for oversampling using:

\begin{equation}
\label{Eqn:PRS_corr}
    Z_{x} = \frac{Z^{\prime}_{x}}{\sigma_{Z^{\prime}_{WN}}}.
\end{equation}

After correction, the error on the corrected PRS \Zx\ is now \sigmaZx$=1$, and a magnitude of $\lvert Z_{x} \rvert > 3$ is considered statistically significant. The number of independent samples can be estimated as

\begin{equation}
    \label{Eqn:n_ind}
    n_{ind} = \frac{n}{\left(\sigma_{Z'_{WN}}\right)^{2}}.
\end{equation}

For the Velocity Dependent HRO analysis, a corrected PRS Z$_{x_i}$ is measured for each integrated intensity map $M(x,y)_{i}$ found at the velocity slab centered at the velocity $v_{i}$. This generates a PRS as a function of velocity.

\section{Results}
\label{Sec:Results}

\subsection{HRO between Band C and Band E}
\label{sec:BandCE_HRO}

\begin{figure}
    \centering  \includegraphics[width=0.5\textwidth]{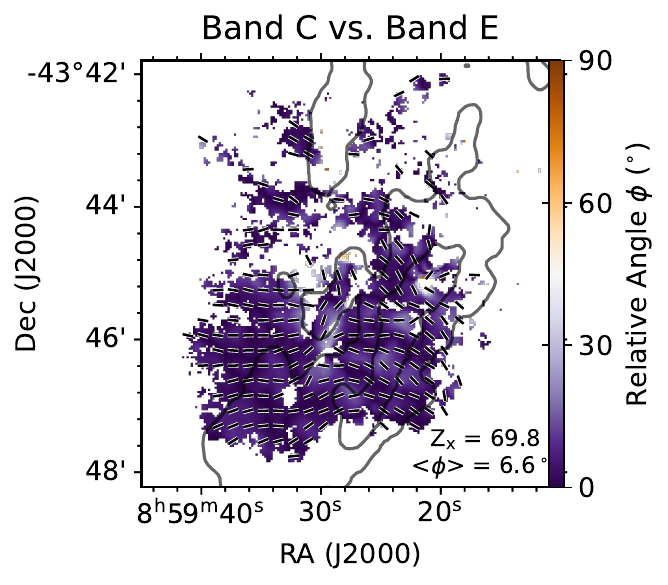}
    \caption{Map of the relative angles $\phi(x,y)$ measured between the magnetic field orientation inferred from HAWC+ Band C (89 $\mu$m) and Band E (214 $\mu$m). The line segments show the plane-of-sky magnetic field orientation inferred from SOFIA HAWC+ Band C polarization data.} 
    \label{fig:BandC_BandE_HRO}  
\end{figure}

In this section, we use the HRO method to compare the magnetic field orientations inferred by SOFIA at 89 $\mu$m (Band C) to 214 $\mu$m (Band E). Figure \ref{fig:BandC_BandE_HRO} shows the relative angles between the Band C ($\hat{B}_{\perp C}$) and Band E ($\hat{B}_{\perp E}$) magnetic field vectors at each pixel, calculated using Equation \ref{Eqn:RelativeAngle}. We find that the relative angles are near parallel $\phi\sim 0^{\circ}$ at most locations in RCW 36, signifying that the magnetic field orientations in the two different bands are highly consistent. This gives a mean relative angle of $\langle\phi\rangle = 6.6^{\circ}$ as well as a a large positive PRS of $Z_x=69.8$ and normalized statistic of $\Tilde{Z}'_{x}=0.95$, indicating a strong preference for parallel relative alignment. A discussion of this result is given in Section \ref{sec:BandCBandE}. 

Since the magnetic field orientations in the two bands are very similar, we present only the Band C HRO results in the main text and defer the Band E results to Appendix \ref{Results_BandE}. The Band C Single Map and Velocity Dependent HRO results are presented in Sections \ref{sec:Results_Single} and \ref{sec:Results_Velocity}, respectively.

 \subsection{Single Map HRO Results}
\label{sec:Results_Single}

Table \ref{tab:PRS} summarizes the Single Map HRO results, where the SOFIA/HAWC+ Band C data has been used to infer the magnetic field orientation. Most tracers have a negative PRS (\Zx), indicating a statistical preference for perpendicular alignment. There are also some notable exceptions that have a positive PRS, indicating a preference for parallel alignment. We discuss the results from the various tracers below.

\begin{table}
    \centering
    \caption{PRS Results for the Single map HRO Analysis using magnetic field orientation inferred from HAWC+ 89 $\mu$m data.}
    \begin{tabular}{ccccccc}
        \hline 
        \vspace{1mm}
        Instrument &
        Map & 
        \Zx \footnote{PRS corrected for oversampling using Eq. \ref{Eqn:PRS_corr}} & \sigmaWN \footnote{The standard deviation for 1000 white noise runs used to correct  oversampling in PRS from Eq. \ref{Eqn:PRS_corr}} & \Zprime \footnote{The uncorrected PRS from Eq. \ref{Eqn:UnCorrPRS}}  & \Znorm \footnote{The normalized uncorrected PRS from Eq. \ref{Eq:normalizedUncorrPRS}} & $n$ \footnote{The number of relative angle points used to calculate the uncorrected PRS.} \\
        \hline \hline
        SPIRE & 500 $\mu$m  
        & -8.8 & 7.9 & -69.5 & -0.48 & 10447 \\
        SPIRE & 350 $\mu$m  
        & -11.2 & 5.7 & -63.9 & -0.44 & 10447 \\
        SPIRE & 250 $\mu$m  
        & -12.4 & 4.2 & -52.2 & -0.36 & 10447 \\
        HAWC+ & 214 $\mu$m 
        & -13.3 & 4.1 & -54.3 & -0.39 & 9798 \\
        PACS & 160 $\mu$m   
        & -9.0 & 4.1 & -37.0 & -0.26 & 10447 \\
        HAWC+ & 89 $\mu$m 
        & -9.3 & 3.7 & -34.7 & -0.27 & 8564 \\
        PACS & 70 $\mu$m   
         & -5.7 & 3.7 & -21.3 & -0.15 & 10447 \\
        IRAC & 4.5 $\mu$m   
        & +5.9 & 4.1 & +24.3 & +0.06 & 85309 \\
        IRAC & 3.6 $\mu$m   
        & +7.7 & 4.1 & +31.1 & +0.08 & 85010 \\
        ALMA ACA & 1.4 mm  
        & -0.6 & 3.5 & -2.2 & -0.03 & 2638 \\
        ALMA 12m & 1.4 mm  
        & -1.0 & 3.8 & -3.8 & -0.01 & 32739 \\
        \emph{Herschel} & N(H$_2$) 
        & -11.0 & 3.9 & -42.7 & -0.30 & 10447 \\
        \emph{Herschel} & Temp 
        & -2.8 & 7.0 & -19.7 & -0.14 & 10447 \\
        LAsMA & $^{12}$CO  
        & -2.2 & 4.7 & -10.3 & -0.07 & 10447 \\
        LAsMA & $^{13}$CO  
        & -6.9 & 4.6 & -32.0 & -0.22 & 10447 \\
        upGREAT & \CII\ 
        & +0.5 & 4.4 & +2.2 & +0.02 & 9071 \\
        upGREAT & \OI\   
        & -3.0 & 7.2 & -21.2 & -0.15 & 9634 \\
         Mopra & HNC  
         & -2.2 & 7.6 & -16.4 & -0.11 & 10447 \\
         Mopra & C$^{18}$O 
         & -2.7 & 7.2 & -19.3 & -0.13 & 10447 \\
         Mopra & N$_{2}$H$^{+}$ 
         & -3.5 & 7.7 & -27.0  & -0.19 & 10447 \\
         \hline
    \end{tabular}
    \label{tab:PRS}
    \begin{minipage}{8.5cm}
    \vspace{0.1cm}
    \small  \textbf{Notes:} The structure map used for each molecular line is a single intensity map integrated over the velocity range specified in Table \ref{tab:Obs} using Equation \ref{Eqn:TotalIntIntensity}. A negative \Zx\ corresponds to an overall preference for perpendicular alignment, while a positive value corresponds to a preference for parallel alignment. The larger the magnitude of \Zx, the stronger the statistical significance of the preferred alignment.
    \end{minipage}
\end{table}

\begin{figure}
    \centering  
    \includegraphics[width=0.47\textwidth]{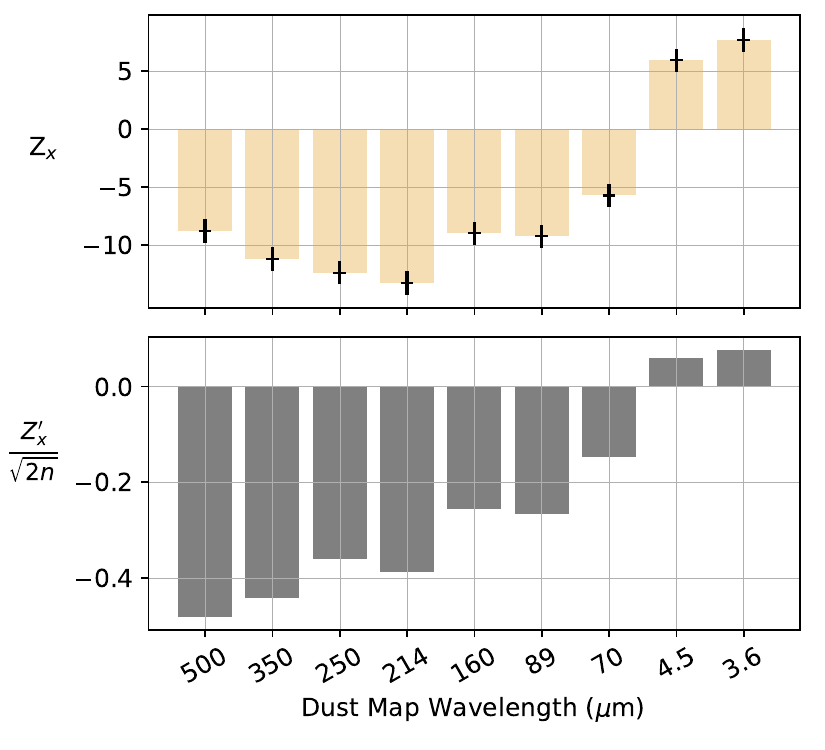}
    \caption{The PRS results from the Single Map HRO analysis which compares the orientation of elongated structures in dust maps of varying observing wavelengths relative to the magnetic field orientation inferred from SOFIA/HAWC+ 89 $\mu$m. The top panel shows the \Zx\ values which have been corrected for oversampling (using Equation \ref{Eqn:PRS_corr}), and indicate the statistical significance of a preference for parallel ($Z_{x} > 0$) or perpendicular alignment ($Z_{x} < 0$) of map structures relative to the magnetic field. The \Zx\ values have errors of 1, as shown by the error bars. The bottom panel shows the normalized uncorrected \Znorm\ values which indicate the degree of alignment. A maximum value of $\Tilde{Z}'_{x}=+1$ corresponds to complete parallel alignment while $\Tilde{Z}'_{x}=-1$ corresponds to complete perpendicular alignment.} 
    \label{fig:PRS_Wavelength}  
\end{figure}

The single-dish dust emission maps show a distinct variation in the magnitude of the PRS values with wavelength. The top panel of Figure \ref{fig:PRS_Wavelength} shows the oversampling-corrected \Zx\ values, which indicate the statistical significance of the PRS, while the bottom panel shows the normalized uncorrected \Znorm\ values, which indicate the degree of alignment. Both panels show that \Zx\ and \Znorm\ are negative for the sub-mm and far-IR \emph{Herschel} and SOFIA data indicating a preference for perpendicular alignment, and positive for the mid-IR \emph{Spitzer} data, indicating a preference for parallel alignment. All the maps have statistically significant values \Zx\ (i.e. $\lvert Z_{x} \rvert > 3$\sigmaZx, where  \sigmaZx$= 1$). A notable trend is seen for the normalized statistic where the \Znorm\ values roughly increase (i.e becomes less negative) as the wavelength decreases, suggesting that successively more structures within the maps align parallel to the magnetic field at shorter wavelengths. In comparison, the trend for the oversampling-corrected statistic \Zx\ decreases from 500--250 $\mu$m and peaks in magnitude at 214 $\mu$m. This is because the oversampling correction factor (\sigmaWN) is proportional to the number of independent samples in the masked area and lower \Zx\ values are expected for the same degree of alignment at longer wavelengths with larger beams, where there are fewer pixels over the same area (see Equation \ref{Eqn:n_ind}). We also ran Monte Carlo simulations to test whether measurement uncertainties in the magnetic field orientation could affect our measured PRS values. We find that the uncertainty in the relative angles $\phi(x,y)$ has a negligible effect on the PRS, resulting in in an uncertainty of $\pm$0.2 for \Zx\ and $\pm$0.002 for \Znorm. These tests and a discussion of our error propagation methods are described in Appendix \ref{sec:PRS_error}.

Figure \ref{fig:PRS_Mom0} summarizes the oversampling-corrected and normalized PRS for the total integrated intensity spectral line maps. Unlike the \Zx\ values for the dust map  shown in Fig. \ref{fig:PRS_Wavelength}, many of the spectral line intensity maps show no overall preference for alignment, or only show a statistically insignificant alignment trend. These maps show different alignment preferences relative to magnetic field in different regions, or overlapping along the line of sight. In Section \ref{sec:Results_Velocity}, we will show that some of these structures that overlap in the integrated intensity map can be decomposed into different line of sight velocity channels. In some cases, particularly for the Mopra observations, the low \Zx\ values are also in part due to lower resolution and higher noise levels of the spectroscopic data. Overall, we note that all gas tracers have a negative \Znorm, signifying a preference for perpendicular alignment, with the exception of \CII\ which has a positive \Znorm.

In Figure \ref{fig:Results_Summary} we identify which structures are aligned with the magnetic field for a select number of datasets. Similar plots for the remaining datasets in Table \ref{tab:PRS} are shown and discussed in Appendix \ref{sec:leftover_single}, Figures \ref{fig:Results_DustApp}--\ref{fig:Results_ALMA}. The right column shows the structure map $M(x,y)$ overlaid with the magnetic field orientation. The middle column shows the relative angle $\phi(x,y)$ calculated at each location in the region, where purple ($\phi\sim0^{\circ}$) is associated with local parallel alignment and orange ($\phi\sim90^{\circ}$) is associated with local perpendicular alignment relative to the magnetic field. The left column summarizes the alignment trend using a histogram of the relative angles.

\begin{figure}
    \centering  
    \includegraphics[width=0.47\textwidth]{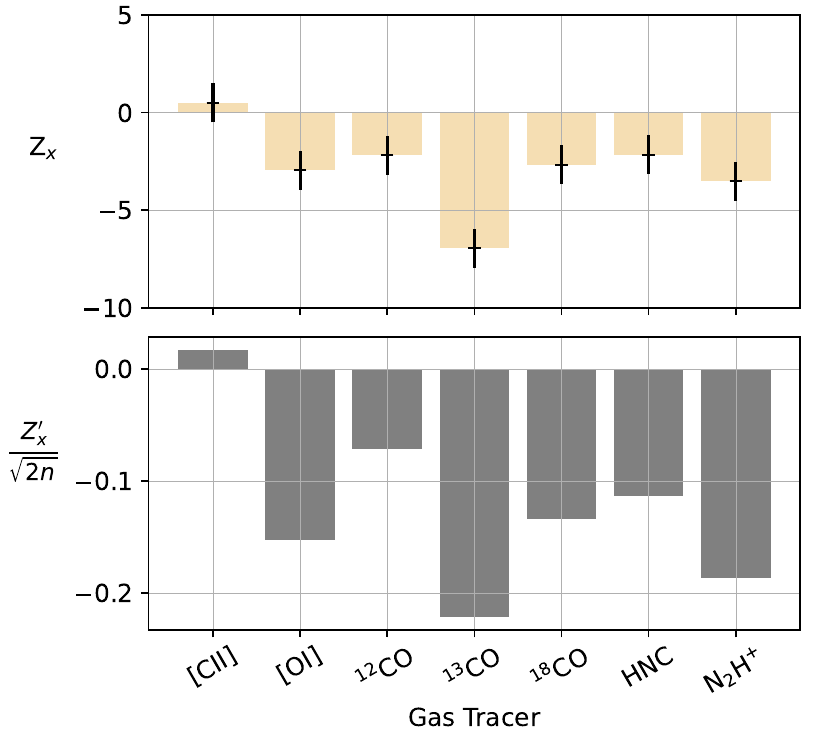}
    \caption{Same as Figure \ref{fig:PRS_Wavelength} but now showing the total integrated intensity spectral line maps.} 
    \label{fig:PRS_Mom0}  
\end{figure}

\begin{figure*}
    \centering  \includegraphics[width=0.86\hsize]{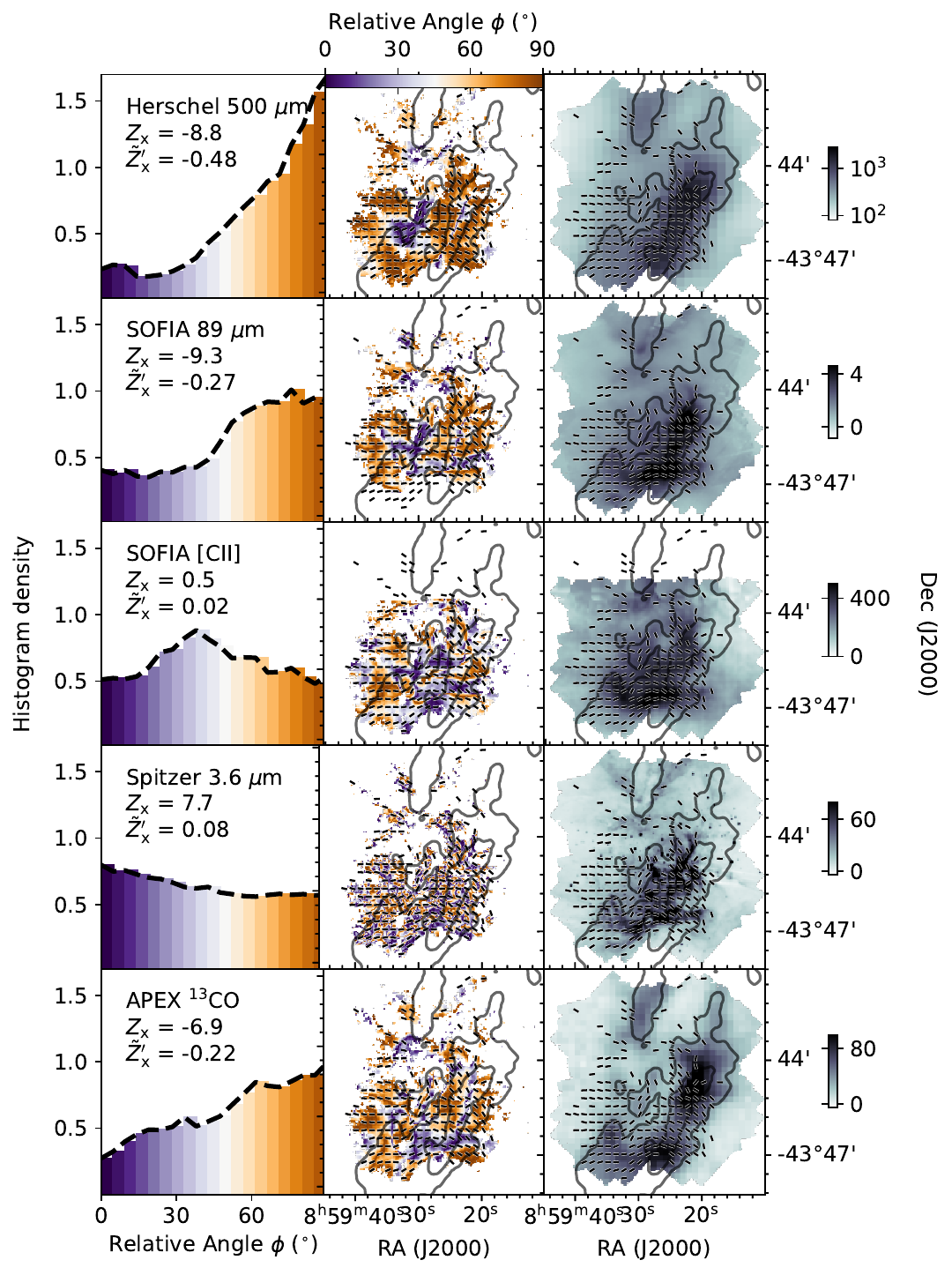}
    \caption{HRO results for selected maps. \textit{Right:} Intensity maps (labeled in the left column text) that are projected onto the HAWC+ 89 $\mu$m grid. The colorbar is units of MJy sr$^{-1}$ for the \textit{Herschel} and \textit{Spitzer} maps, Jy pixel$^{-1}$ for 89 $\mu$m map, and K km/s for the integrated intensity [CII] and $^{13}$CO maps. The \textit{Herschel} 500 $\mu$m and SOFIA 89 $\mu$m maps are in log-scale, while the rest are linear-scale. The vectors show the orientation of the magnetic field inferred from HAWC+ 89 $\mu$m polarized emission, where detections below $3\sigma$ have been  masked. \textit{Middle:} Spatial distribution of relative angles $\phi$(x,y), sharing the same RA/Dec axes as the right column. Only pixels where the inferred magnetic field orientation is not masked have $\phi$(x,y) values. Contours for right and middle panel show column densities of 1.5$\times 10^{22}$ and 4.7$\times 10^{22}$ cm$^{-2}$ (same as Fig \ref{fig:NewData}). \textit{Left:} Histogram density for relative angles in 5$^{\circ}$ wide bins. The color map for angles is the same as the middle panel colorbar, where purple signifies a parallel relative angle of $\phi(x,y)=0^{\circ}$ and orange signifies a perpendicular relative angle of $\phi(x,y)=90^{\circ}$} 
    \label{fig:Results_Summary}  
\end{figure*}

From Fig. \ref{fig:Results_Summary}, we first  compare the relative angle maps for dust emission at wavelengths of 500 $\mu$m and 89 $\mu$m. We note that for both maps, the majority of the relative angles are near-perpendicular and are concentrated at the left and right sides of the dust map, which correspond to the east and west halves of the dense molecular ring labeled in Figure \ref{figure:RGBScale}. This is consistent with the visual observation that the ring is elongated approximately along the north-south direction which is oriented roughly perpendicular to the mostly east-west magnetic field morphology from HAWC+ Band C observations. Both 500 $\mu$m and 89 $\mu$m wavelengths also show near-parallel relative angle measurements ($\phi\sim 0^{\circ}$) within the roughly N-S oriented Flipped-Fil structure, oriented parallel to the local N-S magnetic field. The main difference between the 500 $\mu$m and 89 $\mu$m, however appears to be the south Bent-Fil structure (labeled in Figure \ref{figure:RGBScale}). This structure is traced at the shorter 89 $\mu$m wavelength, but not at 500 $\mu$m. Since the Bent-Fil structures are elongated E-W, parallel to the local magnetic field, this results in the 89 $\mu$m having an overall lower \Znorm\ magnitude, indicating less of a statistical preference for perpendicular alignment relative to the magnetic field. This general trend is noted for all other sub-mm and far-IR the dust maps from 350--70 $\mu$m as well (see Appendix \ref{sec:App_dust}, Fig. \ref{fig:Results_DustApp}), where the emergence of the southern Bent-Fil structure at wavelengths $< 160\,\mu$m results in less negative \Zx\ values as observed in Fig. \ref{fig:PRS_Wavelength}. 

However unlike the far-IR and sub-mm dust maps, the 3.6 $\mu$m \emph{Spitzer} data is less sensitive to the high column density ring structure and instead predominantly traces emission near the north and south Bent-Fils which are oriented parallel to the E-W magnetic field. The lack of perpendicular relative angles from the dense ring results in an overall positive \Zx.

The ALMA continuum maps show no significant preference for either parallel or perpendicular alignment. This is likely because ALMA interferometer is resolving out many of the large-scale dense ring, Main-Fil, Flipped-Fil and Bent-Fils structures (see full discussion in Appendix \ref{sec:App_dust} and Figure \ref{fig:Results_ALMA}). 

Examining the molecular gas maps, we find that $^{13}$CO, which is sensitive to intermediate density gas, is able to trace both the dense molecular ring and the south Bent-Fil structure, resulting in a \Zx\ value comparable to the 89 $\mu$m dust map. 

The \CII\ relative angle map shows that the E-W Bent-Fil structures contribute about the same number of parallel aligned relative orientations ($\phi\sim 0^{\circ}$) as the perpendicular $\phi\sim 90^{\circ}$ relative orientations near the N-S ring. This results in a PRS ($Z_{x}=+0.5$) that is close to 0 and has neither a statistical preference for perpendicular nor parallel alignment relative to the magnetic field. Distinctively, the histogram of \CII\ also does not peak at near-parallel or near-perpendicular angles, but rather close to $\phi\sim 40^{\circ}$. Though it is not statistically significant, the positive \Zx\ result of \CII\ is interesting as it is in contrast with the negative \Zx\ results for all other of all other spectral line tracers, which predominantly trace the molecular dense ring (Fig. \ref{fig:Results_GasApp}). Furthermore, we see that the emission in the integrated intensity \CII\ map correlates with the \emph{Spitzer} emission, which probes warm dust likely found near PDRs. It is therefore noteworthy that the \Zx\ results for both the \CII\ and \emph{Spitzer} maps are positive, indicating that structures associated with PDRs have a preference towards parallel alignment relative to the magnetic field.  

In summary, we find a fairly bimodal trend in the Single Map HRO results. Maps which predominately trace the high-column density ridge and ring structure (such as longer wavelength dust maps and high-density gas tracers) show an overall preference for perpendicular alignment relative to the magnetic field. Whereas maps which trace more diffuse structures near or within the PDR (such as mid-infrared dust maps and \CII) show more of a tendency towards parallel alignment relative to the magnetic field. Maps which show a combination of the two types of structures (such as 160\textendash70 $\mu$m dust maps and low-to-intermediate density gas tracers) show both regions of parallel and perpendicular alignment relative to the local magnetic field, resulting in a final PRS of lower magnitude. We discuss some caveats and considerations of our HRO analysis in Appendix \ref{sec:caveats}. In the next section, we discuss the HRO analysis for different line-of-sight velocity ranges in the spectral line cubes. We use this Velocity Dependent HRO approach to examine the relationship between the orientation of the different line-of-sight gas structures and the magnetic field.
 
\begin{figure}[h]
  \centering
  \includegraphics[width=0.48\textwidth]{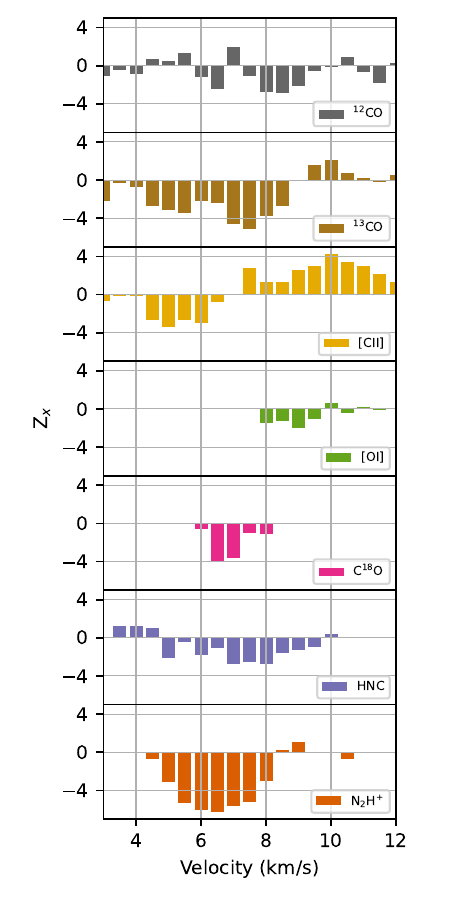}
  \caption{Corrected PRS Z$_{x}$ as a function of velocity for different molecular lines.}
  \label{fig:PRS_Velocity}
\end{figure}

 \begin{figure*}
  \centering
  \includegraphics[width=0.8\textwidth]{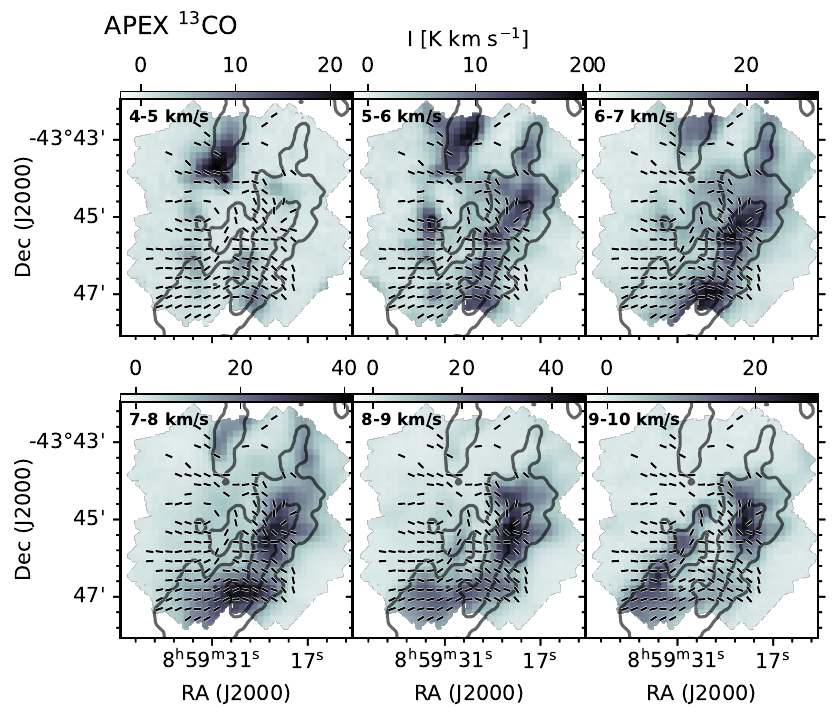}
  \caption{The integrated intensities of $^{13}$CO (3--2) for 1 km/s wide velocity slabs (labeled on the top-left of each panel) from 4--10 km/s. The maps have been projected onto the HAWC+ Band C grid. The color-scale indicates integrated intensity in units of K km s$^{-1}$. The contours are the same as Figure \ref{fig:NewData}. The vectors show the magnetic field orientation inferred from HAWC+ 89 $\mu$m data}
  \label{fig:13CO_Slices}
\end{figure*}

\subsection{Velocity Dependent HRO}
\label{sec:Results_Velocity}

\begin{figure*}
    \centering  \includegraphics[width=0.8\textwidth]{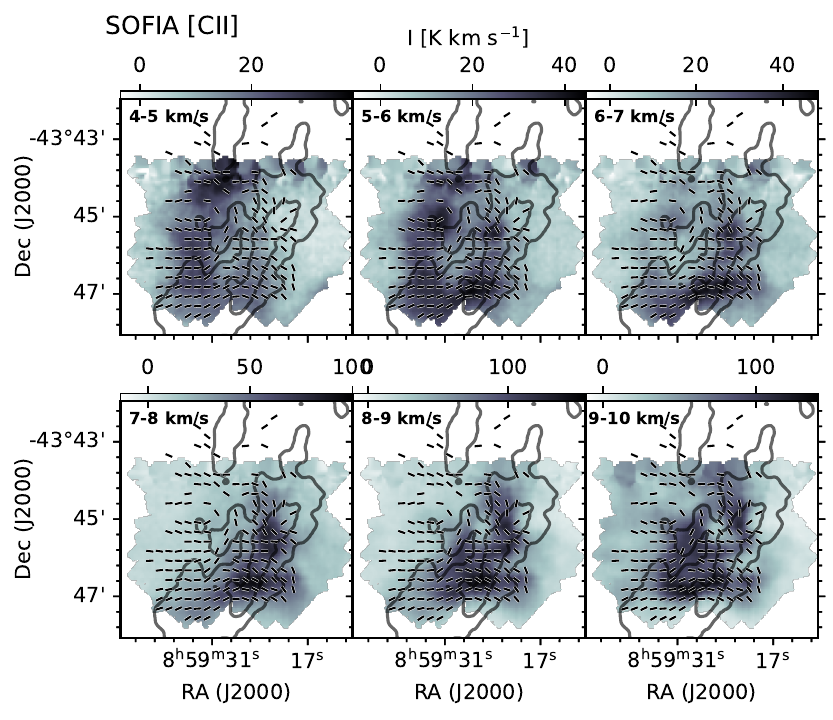}
    \caption{Same as Fig. \ref{fig:13CO_Slices} but for SOFIA \CII\ data.} 
    \label{fig:CII_slices}  
\end{figure*}

\begin{figure*}
    \centering  \includegraphics[width=0.8\textwidth]{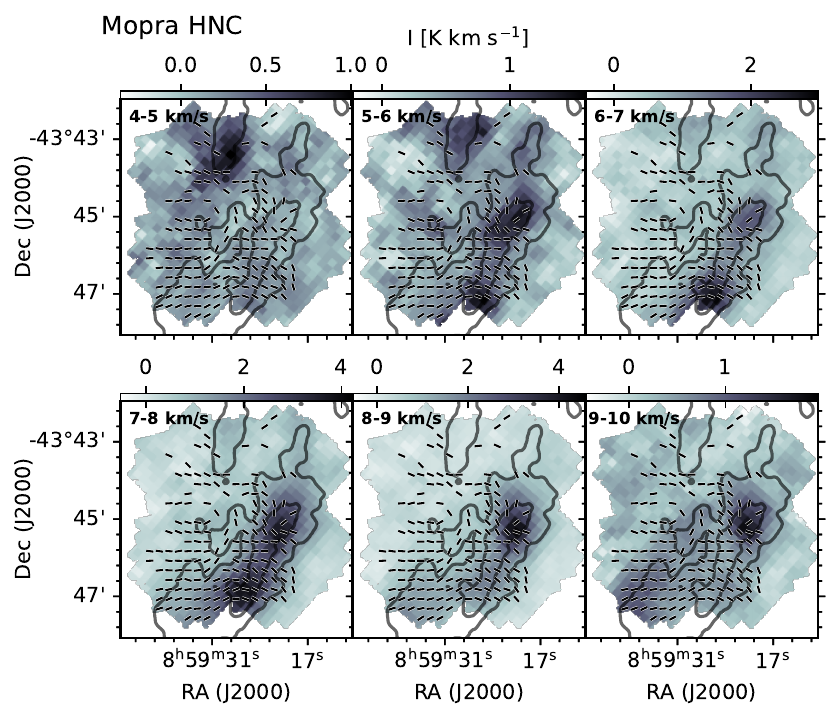}
    \caption{Same as Fig. \ref{fig:13CO_Slices}  
     but for Mopra HNC data.} 
    \label{fig:HNC_slices}  
\end{figure*}

In this section we present the results of the Velocity Dependent HRO analysis which measures the PRS of a spectral cube as a function of line-of-sight velocity, using the method described Section \ref{sec:Methods_velocity}. Figure \ref{fig:PRS_Velocity} shows the corrected PRS results for different spectral lines as a function of velocity. We note that while the magnitude of the \Zx\ is not always statistically significant ($<$ 3-$\sigma$) over the velocity range for all tracers, the overall trend is interesting and consistent with the Single Map HRO results. The intermediate and dense gas tracers, C$^{18}$O, HNC, N$_{2}$H$^{+}$ typically have statistically significant negative \Zx\ values, especially around 5\textendash8 km/s, implying a preference for perpendicular alignment. This velocity range matches the mean line-of-sight velocity of the cloud of around 7 km/s \citep{2013A&A...550A..50M, 2022ApJ...935..171B}. In contrast, the \CII\ PRS results switch from negative \Zx\ values around 5\textendash6 km/s to statistically significant positive \Zx\ values 9\textendash11 km/s, indicating a preference for parallel alignment at higher line-of-sight velocities. 

Figure \ref{fig:PRS_Velocity} also demonstrates the limitations of using a single integrated intensity map for a spectroscopic cube in the HRO analysis, as was done in Section \ref{sec:Results_Single}. Since there can be multiple overlapping elongated structures at different line-of-sight velocities, measuring the PRS from only one integrated intensity map may result in a loss of information on the alignment preferences of kinematically distinct structures. To differentiate which structures are being observed at different velocities, Figures \ref{fig:13CO_Slices}\textendash\ref{fig:HNC_slices} show channel emission maps from 4\textendash10 km/s. 

In the $^{13}$CO channel maps (shown in Fig.  \ref{fig:13CO_Slices}), we notice that emission from the northeastern half of the ring structure is most prominent at $\sim4\textendash6$ km/s, while the southwestern section of the ring is most prominent at $\sim6\textendash8$ km/s. Since the ring including the Main-Fil is oriented N-S, approximately perpendicular to the E-W magnetic field, the overall \Zx\ at these velocities is preferentially perpendicular and therefore negative. However, the area of the northeastern region of the ring is smaller and contains fewer HAWC+ Band C polarization detections, leading to less $\phi(x,y)\sim90^{\circ}$ pixels at 4\textendash5 km/s compared to larger southwestern component of the ring, resulting in a less negative \Zx. At line-of-sight velocities $>$ 8 km/s, the gas traces the Flipped-Fil and north Bent-Fil, causing a weak preference for parallel alignment around 10 km/s.

Similarly, in the \CII\ channel maps (Fig. \ref{fig:CII_slices}) we notice that the eastern half of the ring can be seen in the 4\textendash5 and 5\textendash6 km/s maps, followed by the western half of the ring at 6--8 km/s along with the southern Bent-Fil, all of which result in an overall negative and therefore preferentially perpendicular, negative \Zx\ for these velocity channels. At higher velocities we see emission from the Flipped-Fil and northern Bent-Fil at 8--10 km/s, resulting in an overall positive \Zx\ in these velocity channels. Since the Bent-Fil features are the more prominent in \CII\ than other gas tracers, it has the largest positive \Zx\ magnitude. 

In contrast, HNC (Fig. \ref{fig:HNC_slices}) which is tracing denser gas, mostly shows emission tracing the dense ring (eastern half at 4--6 km/s and western half at 6\textendash8 km/s) has a mostly negative \Zx. The channel maps for the rest of the molecular lines are shown in Appendix \ref{sec:leftover_velocity}, Figures \ref{fig:12CO_slices}\textendash\ref{fig:C18O_slices}, all of which show emission from the dense ring from 4\textendash8 km/s, and the Flipped-Fil and Bent-Fil features from 8\textendash10 km/s.

This switch from negative to positive PRS as a function of line-of-sight velocity is consistent with our Single Map HRO findings. In Section \ref{sec:Results_Single} we noted a bimodal trend in the PRS, where high column density gas tracers showed a preference of perpendicular relative alignment while tracers associated with the PDR and warmer dust showed more parallel relative orientations. From the Velocity Dependent HRO analysis, we now learn that the dense gas and PDR structures are also kinematically distinct, such that the same PRS bimodality is also observed as a function line-of-sight velocity. In the next section, we interpret these results and suggest potential physical mechanisms that may be causing the observed trends.

\section{Discussion}
\label{sec:Discussion}

The goal of this section is to better understand the physical processes behind the observed magnetic field geometry and morphology of the star-forming structures within the RCW 36 region. We are particularly interested in understanding the energetic impact of the magnetic field and stellar feedback in shaping the gas dynamics. To do this, we examine the magnetic field observations inferred from HAWC+ in Section \ref{sec:MagneticModel}, followed by an interpretation of the HRO results and discussion of the origins of the Flipped-Fil structure in Sections \ref{sec:Interpret_HRO} and \ref{sec:FlippedFil_origin}, and finally comment of the energetic balance of the region in Section \ref{sec:EnergeticBalance}.

\subsection{The Magnetic Field Structure of RCW 36}
\label{sec:MagneticModel}

In this section we discuss the polarization data from SOFIA/HAWC+ in more detail to try and infer the density scales for which the RCW 36 magnetic field is being traced i.e., the  population of the dust grains which contribute to the majority of the polarized emission. To do this, we first estimate the optical depth of the dust emission in Section \ref{sec:OpticalDepth} and compare the polarization data and magnetic field morphologies at the different HAWC+ wavelengths in Section \ref{sec:BandCBandE}.

\subsubsection{Optical Depth of Dust Emission}
\label{sec:OpticalDepth}
To better understand the location of the dust grains contributing to SOFIA/HAWC+ polarized emission maps, we estimate the optical depth $\tau_{\nu}$ to check whether the magnetic field is being inferred from the average column of material along the entire line of sight or if it is tracing only the outer surface layer of an opaque dust cloud. The full method is discussed in Appendix \ref{App:OptDept} and summarized here. We use $\label{eqn:tau} \tau_{\nu} = N_{H_2} \mu m_{H} R_{dg} \kappa_{\nu}$, where $N_{H_2}$ is the column density, $\mu$ is the mean molecular weight, $m_{H}$ is the mass of Hydrogen, $R_{dg}$ is the dust-to-gas ratio, and $\kappa_{\nu}$ is the dust opacity. We adopt the same dust opacity law (given in Appendix \ref{App:OptDept}) as in previous HOBYS and \textit{Herschel} Gould Belt Survey \citep[HGBS:][]{2010A&A...518L.102A} studies \citep[e.g.][]{2010A&A...518L..77M, 2014A&A...562A.138R}. The opacity law is independent of temperature and assumes a dust-to-gas fraction of 1\%. We use a \emph{Herschel} column density map derived by \cite{2011A&A...533A..94H} which has angular resolution of 36\arcsec{}, and is different than the 18\arcsec{} column density map listed in Table \ref{tab:Obs} used for the HRO analysis. We choose to use the 36\arcsec{} column density map since the resolution matches the temperature map. The assumed dust opacity law from \cite{1983QJRAS..24..267H} and spectral index is also consistent with \cite{2011A&A...533A..94H} and \cite{2014A&A...562A.138R}. 

We estimate that the optical depth for 214 $\mu$m (Band E) is optically thin $\tau_{\nu}\ll 1$ everywhere within RCW 36 (see Fig. \ref{fig:OptDept} in Appendix \ref{App:OptDept}). At 89 $\mu$m, we also find that the emission is fairly optically thin $\tau_{\nu} < 1$ at 89 $\mu$m (Band C) for most regions, except for certain locations within the Main-Fil where the $\tau_{\nu}$ can reach values of $\sim$1.4. It should be noted that these optical depth estimates are uncertain due to the difference in resolution and the possibility of emission from very small dust grains (see Appendix \ref{App:OptDept} for details). As a first approximation however, we find that for most regions in RCW 36 the dust emission should be optically thin in HAWC+ Band C and E, and we should therefore be able to probe the full dust column. 

\subsubsection{Magnetic Field Comparison}
\label{sec:BandCBandE}

\begin{figure}
    \centering  \includegraphics[width=\hsize]{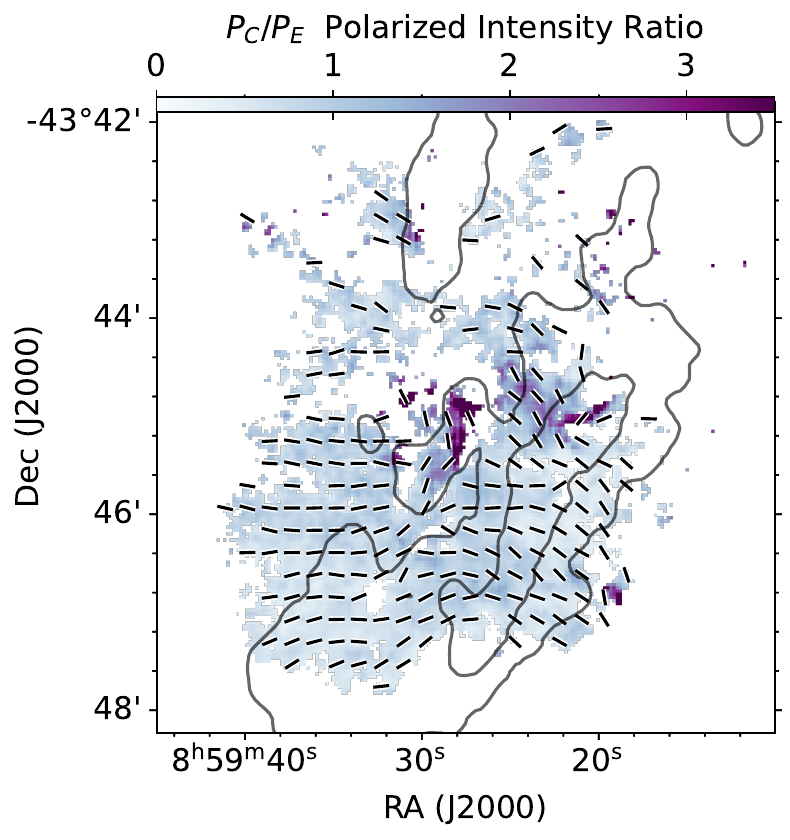}
    \caption{The total polarized intensity measured for HAWC+ Band C (89 $\mu$m) divided by the total polarized intensity measured for HAWC+ Band E (214 $\mu$m). The ratio of intensity $P_{C}/P_{E}$ is shown by the colorbar, where a ratio $\sim$1 corresponds to where the intensities are roughly equal. The Band E polarized intensity was projected on to the Band C grid, and the Band C data was smoothed to the Band E resolution. The Band C polarized emission detections below a 3$\sigma$ signal-to-noise cutoff have been masked.} 
    \label{fig:BandC_ERatio}  
\end{figure}

In this section, we discuss the wavelength dependence of the polarization data from HAWC+ since the Band C (89 $\mu$m) and Band E (214 $\mu$m) may be sensitive to different dust grain populations. Emission at 89 $\mu$m is typically more sensitive to warm (T $\geq$ 25 K) dust grains and less sensitive to cold (T $\leq$ 15 K) dust grains. This is in contrast to 214 $\mu$m, which can also probe the magnetic field orientation in colder, more shielded dust columns. However, in Section \ref{sec:BandCE_HRO}, we used the HRO method to statistically show that the magnetic field morphologies inferred from Band C and Band E are almost identical.

This high degree of similarity could suggest that the Band E observations may be measuring polarized radiation mostly emitted by warm dust grains, similar to Band C. Alternatively the magnetic field morphology in regions where the dust grains are warmer (T $>$ 25 K) may be similar to the field morphology over a wider range of dust grain temperatures. In either case, the Band C polarization data is tracing dust grains with higher temperatures which, in a high-mass star-forming region like RCW 36, are likely being heated from the radiation of the massive stellar cluster. This warm dust is therefore probably located near the \HII\ expanding gas shell and associated PDR. This is also supported by the observation that dust polarized intensity of Band C appears to correlate with the PDR-tracing \CII\, and anti-correlate with the ALMA ACA continuum which traces cold dense cores (as shown in Appendix \ref{sec:CorrelateTracers}). Therefore the HAWC+ magnetic field is likely weighted towards the surface of the cloud near the PDR, rather than the colder denser dust structures.

Aside from dust temperature, the similarity between Band C and Band E magnetic field orientations may further indicate that the magnetic fields are likely being traced at comparable scales and densities in the two bands. Moreover, a consistent magnetic field morphology can be expected at the different wavelengths if the dust emission is optically thin, as previously suggested. 

One noteworthy difference between the Band C and Band E datasets, however, was found by comparing the total polarized intensity given by Eq. \ref{Pol_intensity} in Band C  ($P_{C}$; smoothed to the resolution of Band E) to Band E ($P_{E}$). Figure \ref{fig:BandC_ERatio} shows that the ratio of $P_{C}/P_{E}$ is close to unity for the majority of RCW 36 except for certain regions. These regions have higher polarized intensities in the Band C map than they do in Band E by a factor of $\sim2\textendash4$. Interestingly, the regions also overlap with where the HAWC+ magnetic field is seen to deviate from the general E-W trend of the BLASTPol magnetic field, i.e the Flipped-Fil and north Bent-Fil. This may be because the dust grains traced by the Band C map produce radiation with a higher degree of linear polarization, due to higher grain alignment efficiency based on a change in temperature and/or emissivity. A similar analysis has been performed by \cite{2008ApJ...679L..25V} and \cite{2019ApJ...882..113S} who also compared the polarization ratio in different bands. Radiative alignment torques, which are thought to be responsible for the net alignment of the dust grains with their short axes parallel to the magnetic field, require anisotropic radiation fields from photons of wavelengths comparable or less than the grain size \citep{2015ARA&A..53..501A}. In this case, we may expect to see the polarization efficiency increase towards regions where the dust has been heated by the young star cluster, such as the PDR as was noted for the Bent-Fils. Another possibility is that the magnetic field lines are more ordered in the gas traced by the warm dust which is being preferentially traced in Band C. More ordered fields could mean less cancellation of the polarized emission and therefore a higher polarized intensity in comparison to a sight-line with more tangled fields. The geometry of the region is also a consideration. The warm dust structures could be inclined at a different angle compared to the cooler layers, as the dust polarized emission is only sensitive to the plane-of-sky magnetic field component. 

\subsection{Interpretation of HRO Results}
\label{sec:Interpret_HRO}
\subsubsection{Preferentially Perpendicular Alignment for Dense Tracers}
\label{sec:Discussion_Perp}
In Section \ref{sec:Results_Single}, we found that the structure maps $M(x,y)$ which predominantly trace dense structures such as the ring and Main-Fil showed a statistically significant preference for perpendicular alignment relative to the filament-scale (7.8\arcsec{} FWHM) magnetic field probed by HAWC+ Band C. This result is consistent with previous large-scale HRO studies which compared the alignment of structures within the Central-Ridge of Vela C relative to the cloud-scale magnetic field probed by BLASTPol at 250, 350 and 500 $\mu$m (2.5--3\arcmin{} FWHM) \citep{2017A&A...603A..64S, 2019ApJ...878..110F}. These studies found that the relative alignment between large-scale structures in the Vela C cloud and the magnetic field is column density and density dependent.  

\cite{2017A&A...603A..64S} showed that for both the entire Vela C cloud and the Centre-Ridge region (containing RCW 36), the relative alignment trend transitions from preferentially parallel or no preferential alignment at low column densities, to preferentially perpendicular at high column densities. They find that the transition occurs at a threshold column density of $N_{H}\sim 10^{21.8}$. Additionally, \cite{2019ApJ...878..110F} compared the orientation of the magnetic field inferred from BLASTPol 500 $\mu$m to integrated line intensity maps of different molecular lines tracing low, intermediate and high density structures averaged over the entire Vela C Cloud. They found that the low density gas tracers were more likely to align parallel to the magnetic field while intermediate to high density tracers were more likely to align perpendicular, with the transition occurring at a density of $n_{\mathrm{H_{2}}} \sim 10^{3}$ cm$^{-3}$. This signature transition to preferentially perpendicular alignment at a critical column density has been observed for several other molecular clouds as well \citep[e.g.,][at 10\arcmin{} FWHM]{2016A&A...586A.138P}.

In this work, we do not see this transition as a function of column density and only observe a preference for perpendicular relative alignment for different column density bins. This could be because RCW 36 region is the densest region within Vela C (with $N_{H} \gtrsim 10^{22.4}$) and most of its structures are above the critical column density.  

We also compare our work to magnetic field studies done on comparable small-scales. \cite{2023ApJ...948..109K} find that the Musca filament is oriented roughly perpendicular to the surrounding magnetic field morphology, as traced by SOFIA/HAWC+ 214 $\mu$m observations. Moreover, \cite{2021ApJ...918...39L} applied the HRO method to dense cores in the Ophiuchus molecular cloud and found similar results of preferential perpendicular alignment between high column density, elongated filament and core-scale structures in $\rho$ Oph A and $\rho$ Oph E relative to the magnetic field traced by SOFIA/HAWC+ 154 $\mu$m observations. The prevalence of this perpendicular relative alignment trend across different star-forming regions in varying molecular cloud environments suggests that shared physical processes may be underlying the observations.

Possible interpretations of such processes have been explored by comparing observations to simulations. For instance, \cite{2013ApJ...774..128S} propose that the degree of magnetization of a cloud impacts the trend in relative alignment, where the high-magnetization case specifically reproduces the transition from preferentially parallel to preferentially perpendicular at a critical density. Other studies such as \cite{2016ApJ...829...84C} reason that the preferentially parallel relative alignment occurs in magnetically dominated (sub-Alfvénic) gas, while preferentially perpendicular relative alignment occurs in turbulence dominated (super-Alfvénic) gas with the transition occurring when the kinetic energy due to gravitational contraction becomes larger than the magnetic energy. This connection to the energy balance is consistent with \cite{2017A&A...607A...2S}, who demonstrate that a transition from parallel to perpendicular relative alignment can occur as a result of convergent velocity flows, which could be due to gravitational collapse. They also find that the transition in alignment can occur when the large-scale magnetic field is strong enough to impose an anisotropic velocity field and set an energetically preferred direction in the gas flow. However, simulations also caution that projection effects are an important consideration in the interpretation of HRO results as \cite{2020MNRAS.497.4196S} showed that the relative orientation trends also strongly dependent on viewing angle.

Based on these studies, we propose that the large-scale magnetic field surrounding RCW 36 may have been dynamically important during it's formation, allowing gas to flow preferentially parallel to the E-W magnetic lines. This may have resulted in the formation of an elongated molecular gas sheet or filament (currently the Centre-Ridge) since material could have been inhibited in collapsing perpendicular to the magnetic field lines in the N-S direction. As the region went on to form stars, \cite{2013A&A...550A..50M} suggest that ionizing radiation from the massive star cluster would have then reshaped the surrounding gas into a bipolar nebula, forming a ring of dense material at the center as an \HII\ region expanded into the elongated structure. Both the BLASTPol and HAWC+ maps show that the magnetic field lines pinch near the waist of the bipolar nebula which could be evidence that the ram pressure may be overpowering the local magnetic pressure in that region, as the magnetic field lines are being warped along with the flux-frozen gas. So while the magnetic field may have set a preferred direction of gas flow during the formation of Centre-Ridge and Main-Fil, it may no longer be energetically significant across all of the RCW 36 region since the birth of the massive stars.

\subsubsection{Parallel Alignment for PDR Tracers}
\label{sec:Discuss_parallel}
Section \ref{sec:Results_Single} also showed that some regions and tracers had a preferential parallel alignment between elongated structures and the inferred magnetic field. The decrease in the statistic magnitude $\lvert$\Zx $\rvert$ from dust map wavelengths shorter than 214 $\mu$m, was found to be largely due to the gradual emergence of the north and south Bent-Fil features (labeled in Fig. \ref{figure:RGBScale}), which are elongated along the orientation of the HAWC+ Band C magnetic field lines. The emergence of Bent-Fils towards shorter wavelength  (70\textendash214 $\mu$m) \emph{Herschel} and SOFIA dust maps, implies that these features are likely tracing cloud structures with warmer dust populations, near the PDR. The north and south Bent-Fils are also traced by the \emph{Spitzer} mid-infrared 3.6\textendash4.5 $\mu$m maps, which are sensitive to emission from hot dust found near the PDR. 

The observation of a preferential parallel relative alignment between the direction of elongation of the Bent-Fils and the local Band C magnetic field orientation can be explained by the coupling of the gas and the magnetic field. We propose that the stellar feedback in the form of ionizing radiation from the high-mass cluster may be warping the flux-frozen gas, thereby dragging the magnetic field lines along with it. The higher resolution and/or shorter wavelengths of the SOFIA/HAWC+ observations are able to trace the regions where the magnetic field orientation appears to be altered from the otherwise uniform east-west geometry traced by 500 $\mu$m BLASTPol observations. The altered field lines follow the warped morphology seen for the bright-rimmed Bent-Fil regions traced by hot dust and \CII\ and \OI\ PDR tracers. 

Furthermore, the Velocity Dependent HRO results (see Section \ref{sec:Results_Velocity}) show that the Flipped-Fil and Bent-Fil structures had a line-of-sight velocity of $\sim$8\textendash10 km/s, while the ring and Main-Fil structures were seen at velocities of $\sim$5\textendash7km/s. If the Bent-Fil features are in fact being warped by expansion pressures from the ionization front, then it may be expected that these features have different velocities as compared to the dense structures which may be more shielded. \cite{2022ApJ...935..171B} estimate an expansion velocity of 1\textendash2 km/s for the dense molecular ring and and \CII\ expanding shells in the bipolar cavities with velocities of $\sim$5 km/s. However, expansion is only one explanation and there are others plausible reasons as to why dense ring and PDR regions have different line-of-sight velocities such as rotations, tidal forces, etc. If the magnetic field lines are indeed being altered by the radiation from the massive stars, then this may suggest that the magnetic field pressure may not be sufficient to the support the cloud structures against the kinetic energy injected by stellar feedback. 

While the Flipped-Fil also shows a strong preference for parallel alignment relative to the local magnetic field and similar line-of-sight velocities as the Bent-Fils, it is not as clear at this stage whether the Flipped-Fil is an irradiated structure associated with warped gas near the PDR. Unlike the Bent-Fils, the Flipped-Fil is not preferentially observed at shorter wavelength dust maps but rather, appears faintly in dust emission across the wavelengths 500\textendash70 $\mu$m (see Fig. \ref{fig:Results_Summary} and Appendix \ref{sec:App_dust}, Fig. \ref{fig:Results_DustApp}). Furthermore, the Flipped-Fil is not traced by the \emph{Spitzer} maps (see Fig. \ref{fig:Results_Summary} and Appendix \ref{sec:App_dust}, Fig. \ref{fig:Results_Leftover}), which may be expected if the structure was associated with warmer dust grains. A full discussion of the origins of the Flipped-Fil is presented in the next section. 

\subsection{Origins of the Flipped-Fil}
\label{sec:FlippedFil_origin}
One region of particular interest throughout this study has been the Flipped-Fil (labeled in Fig. \ref{figure:RGBScale}) due to the N-S orientation of the magnetic field lines locally within the filament, which is in stark contrast with the general E-W orientation of the surrounding HAWC+ Band C magnetic field morphology. While the magnetic field lines appear to deviate slightly from the E-W trend in several regions such as that Bent-Fils, the Flipped-Fil region is the most striking feature as the magnetic field lines appear to change direction more abruptly and are almost orthogonal to the magnetic field of the surroundings. 

Observational effects like projection may be contributing to the abrupt 90$^{\circ}$ change in 2-D orientation, which may not be as drastic in 3-D. A change in the grain alignment mechanism of the dust grains could also cause the near-discontinuous behaviour of the Flipped-Fil if the reference direction for grain alignment changed from the magnetic field to the radiation field, as has been theorized for other high-mass star forming regions such as the Orion Bar \citep{2023ApJ...951...97L}. 

The change in magnetic field orientation within the Flipped-Fil can also be explained through physical origins. One plausible formation scenario was presented by \cite{2018ApJ...860L...6P} studying the magnetic field morphology of the Pillars of Creation seen in M16, which resembles the morphology of the Flipped-Fil. The scenario \citep[detailed schematically in Figure 5 of][]{2018ApJ...860L...6P} is summarized here. An ionization front fueled by photon flux from a massive radiating star or cluster is envisioned to approach molecular gas, which may have regions of varying density. The gas being dragged by the ionization front may bow-around the an over-density to form an elongated pillar. The flux-frozen magnetic field lines within the pillar would then follow the gas motion and end up perpendicular relative to the background magnetic field orientation. Such a structure could remain stable as the compressed magnetic field lines would provide support against radial collapse since gas flow perpendicular to the field lines would be inhibited. The pillar may gradually erode in the length-wise direction however, as gas flow parallel to the field lines would still be allowed. 

While such a physical model may be applicable to a structure similar to the Flipped-Fil, there are obvious differences between our observations and the Pillars of Creation. In spectral line data of the Flipped-Fil is observed at line-of-sight velocities of $\sim$8\textendash10 km/s, which is red-shifted compared to the northern and southern halves of the dense ring. This arrangement could have occurred if the expansion of the \HII\ region swept up the Flipped-Fil and pushed it behind the massive cluster such that it is currently at a further distance away from us and thus receding at a faster line-of-sight velocity than the main ridge. It is also difficult to distinguish from a 2-D projection on the plane-of-sky if the Flipped-Fil is indeed a pillar and column-like structure or whether it is a ridge of material. Moreover, while The Pillars of Creation are photoionized columns. It is not immediately obvious if the Flipped-Fil is directly associated with the PDR as it is not seen in the \emph{Spitzer} maps which trace hot dust, but is seen in the \CII\ and \OI\ integrated intensity maps, which traces irradiated dense gas. The lack of mid-infrared emission towards the Flipped-Fil is likely not due to absorption from foreground structures as the region is associated with \Halpha\ emission (see Fig. \ref{figure:RGBScale}). Additionally, at a column density of \Av\ $\sim 13$, the Flipped-Fil is traced by low and intermediate gas tracers such as $^{12}$CO and $^{13}$CO indicating that it is has molecular gas component, but is not quite dense or cold enough to be traced as clearly in N$_{2}$H$^{+}$ and HNC.  

The Flipped-Fil thus shows clear differences from the Bent-Fils which are likely associated with the warm dust structures and traced by shorter wavelength maps (3.5\textendash160 $\mu$m) as well as PDR tracers such as \CII\ and \OI. The Pillars of Creation formation scenario suggested for the Flipped-Fil can also be applicable to the Bent-Fils. In this picture, the star-forming clumps seen in ALMA ACA continuum data (see Fig. \ref{figure:RGBScale}) could be the over-dense structures envisioned Figure 5 of \cite{2018ApJ...860L...6P}, around which the bright rimmed Bent-Fil structures are being bowed around. The orientation of the bow-shapes then may suggest the direction of these ionization fronts. This model is similar to \cite{2013A&A...550A..50M} who suggest from comparisons of numerical simulations by \cite{2012A&A...546A..33T, 2012A&A...538A..31T} that the bright rims or what we call `Bent-Fils' are the result of density enhancements in thin shells due to gas compression around the pillar-like structures.

While this pillar formation and similar origin scenarios for the Flipped-Fil and Bent-Fils are certainly plausible, there is insufficient evidence for it to be the favoured explanation. Higher resolution infrared observations may help distinguish these structures better, giving more insight to their morphology and origin. 

\subsection{Energetic Balance}
\label{sec:EnergeticBalance}
In this section we examine the energetic balance of the RCW 36 region in light of the new HAWC+ polarization observations presented in this work.  

The suggestion that the flux-frozen magnetic field lines are transitioning from their mostly E-W cloud-scale geometry to align parallel with feedback-associated structures implies that the magnetic field pressure must be less than the ram pressure. This change in morphology indicates that the magnetic field is being altered by feedback as it is unable to support the cloud structures from warping. Setting the ram pressure in equilibrium with the magnetic field pressure would therefore give an upper-limit on the magnetic field strength. \cite{2022ApJ...935..171B} estimate the ram pressure energy density within the ring (labeled in Fig. \ref{figure:RGBScale}) to be $u_{ram}$ = $0.41$\textendash$3.67\times 10^{-10}$ erg cm$^{-3}$. Assuming equipartition, we set the ram pressure energy density $u_{ram}$ equal to the magnetic energy density $u_{B}$ so that:
\begin{equation}
    u_{ram} \geq u_{B} = \frac{B^{2}}{8\pi} \mathrm{(cgs)},
\end{equation}

and the the upper-limit on the magnetic field strength is estimated to be $B=33$\textendash$99\,\mu$G. This is lower than the magnetic field strength of 120 $\mu$G estimated by \cite{2016ApJ...830L..23K} for the Centre-Ridge using the Davis-Chandrasekhar–Fermi method. Furthermore, our estimate is also lower in comparison to similar high-mass star-forming regions. For instance, DR 21 is measured to have a magnetic field strength of 130 $\mu$G \citep{2021MNRAS.501.4825K} and RCW 120 is estimated to have 100 $\mu$G \citep{2022RAA....22g5017C}. Since our upper-limit is crude and based on the assumption that the feedback is ram pressure dominated, we may be underestimating the magnetic field strength. 

\cite{2022ApJ...935..171B} also calculate that the turbulent energy density within the ring is $u_{turb}= 4.1$\textendash$5.1\times 10^{-10}$ erg cm$^{-3}$ which is comparable to the ram pressure energy density and our estimated upper-limit for the inferred magnetic field energy density. However, the magnetic field energy is likely not much weaker than the turbulent energy since a fairly ordered (rather than tangled) magnetic field geometry is observed in RCW 36, which is a signature of sub- or trans-Alfvénic conditions (where $u_{B} \geq u_{turb}$) \citep{1998ApJ...508L..99S}. If the turbulent energy was dominant, we would expect the magnetic field orientation to have more random variations, which would decrease any alignment trend and result in \Zx\ values with smaller magnitudes than our current measurements. Alternatively, the effects of turbulence on the magnetic field morphology may not be visible on the spatial scales probed by SOFIA/HAWC+ if the size of the turbulent eddies are smaller than the size of the beam, such that the polarization component from turbulent motion cancels out along the line-of-sight to give the appearance of low dispersion, as demonstrated in \cite{1989ApJ...346..728J}. On filament-scales, the ordered magnetic field observations from HAWC+ suggest a near-equipartition between the magnetic field energy and turbulent energy, with the ram pressure from stellar feedback dominating in certain regions.

This interpretation is different from the large-scale HRO results using BLASTPol, which suggested that the magnetic field may have been dominating the energetic balance, setting a preferred direction of gas motion on cloud scales for the dense Centre-Ridge to form preferentially perpendicular relative to the magnetic field (see Section \ref{sec:Discussion_Perp}). This indicates that the dynamic importance of the magnetic field may be scale dependent in this region or that the energetic balance has changed since the formation of the original generation of stars which is currently driving the feedback within RCW 36. It should also be noted that, since the filament-scale magnetic field traced by HAWC+ Band C is weighted towards warm dust, the magnetic field may only be less dynamically important near the PDR. Whether this is the case within the cold dense star-forming clumps remains unclear. Presumably gravitational in-fall will also be a strong contributing force to the energetic budget on core scales. A more in-depth analysis of the energetic balance within the cores and clumps, polarization data at longer wavelengths and higher resolutions, such as the polarization mosaics from our ALMA 12-m Band 6 program, is needed. We leave the analysis of that dataset for future work. 

\section{Conclusion}
\label{sec:Conclusion}
The motivation of this work was to better understand the combined influence of stellar feedback and magnetic fields on high-mass star formation. To do this, we targeted the extensively studied region RCW 36 in the Vela C giant molecular cloud, which has been previously observed using many different tracers. Adding to this suite of complimentary data, we presented new, higher resolution observations of the magnetic field morphology inferred from SOFIA/HAWC+ linearly polarized dust emission maps at 89 and 214 $\mu$m at filament-scales as well as ALMA Band 6 continuum 1.1-1.4 mm data at clump scales.

We then employed the Histogram of Relative Orientations (HRO) method to compare the orientation of the HAWC+ magnetic field to the orientation of physical structures in RCW 36 as traced by 7 spectral lines and dust emission and continuum maps ranging from 3.6 $\mu$m to 1.4 mm, for a multi-scale, multi-wavelength study. Comparing our HRO results to previous larger cloud-scale studies and simulations, we discussed the implications of our findings on the energetic importance of the magnetic field in RCW 36. The main conclusions of this analysis are:

\begin{enumerate}
    \item We find that the inferred filament-scale magnetic field from HAWC+ generally matches the east-west morphology of the cloud-scale magnetic field inferred from BLASTPol, except for a few notable regions of interest. One exception we call the Flipped-Fil region, where the field switches to a roughly north-south orientation and the other exception we call the Bent-Fils region, where the field follows a bent shape around star forming clumps. We also find that the magnetic field morphologies inferred by Band C (89 $\mu$m) and Band E (214 $\mu$m) are highly similar, indicating that they may be tracing similar dust grain populations, scales and densities.
    
    \item The HRO analysis between the inferred magnetic field and single intensity maps show differences in orientation between dense gas tracers and PDR tracers. Structures observed in dense gas tracers show a preference for perpendicular alignment relative to the magnetic field, whereas the tracers of warm dust and the PDR show a preference for parallel relative alignment. The aforementioned Flipped-Fil region, however, tends to be preferentially parallel in most tracers for which it is well detected, indicating that this is a special case.
    
    \item Repeating the HRO analysis for different of line-of-sight velocities in the spectroscopic data cubes shows that the relative alignment of structures also varies with velocity. Structures associated with dense gas show a preference for perpendicular alignment relative to the magnetic field at line-of-sight velocities of 4--7 km/s, while structures associated with the PDR show a preference for parallel alignment at velocities of 8--11 km/s. This technique allows us to disentangle otherwise overlapping structures in the single integrated intensity map.  
    
    \item The finding the dense ridge of RCW 36 is oriented perpendicular to the magnetic field is consistent with previous cloud-scale HRO studies of the Centre-Ridge within Vela C \citep{2017A&A...603A..64S, 2019ApJ...878..110F}. Comparing this result to studies which applied the HRO method to synthetic observations of MHD simulations \citep[e.g.,][]{2016ApJ...829...84C, 2020MNRAS.499.4785K} suggests that the magnetic field may have been dynamically important on cloud scales when the dense ridge of RCW 36 region first formed, however this may no longer be the case after the formation of the massive stars.     
    
    \item The HRO results from the warm dust and PDR tracers suggest that the magnetic field lines are perhaps being altered near the ionization front such that they align parallel relative to gas warped by stellar feedback. This could indicate that ram pressure and radiation from the nearby massive cluster may be dominating the energetic balance on filament-scales. This is potentially causing the flux-frozen magnetic fields to be bent in directions which follow the elongation of the bright-rimmed Bent-fil structures. The parallel relative alignment observed for the Flipped-Fil may have resulted from a formation scenario similar to what has been suggested by \cite{2018ApJ...860L...6P} for the Pillars of Creation where gas bows around an over-density creating a pillar-like structure in which the local magnetic field is rotated orthogonally in comparison to the background magnetic field orientation. 
\end{enumerate}

In conclusion, the SOFIA/HAWC+ polarization data provided new insights into the RCW 36 region, particularly regarding how the magnetic field may have been altered near the PDR region due to ionization from the massive stellar cluster. The filament-scale HRO analysis highlighted structures showing parallel alignment relative to the local magnetic field which were not observed in previous HRO cloud-scale studies. This altered magnetic field near the PDR may impact the formation of next-generation stars by influencing gas dynamics. Thus comparing the magnetic field from higher resolution, shorter wavelength polarization data to PDR tracers may offer useful insight when studying the impact of feedback on the magnetic field in other high mass star-forming regions as well.

\section*{Acknowledgments}
We thank the referee for their discerning feedback which has greatly improved the presentation of this work. This research has made use of data from the \emph{Herschel} Gould Belt survey (HGBS) project (http://gouldbelt-herschel.cea.fr). The HGBS is a \emph{Herschel} Key Programme jointly carried out by SPIRE Specialist Astronomy Group 3 (SAG 3), scientists of several institutes in the PACS Consortium (CEA Saclay, INAF-IFSI Rome and INAF-Arcetri, KU Leuven, MPIA Heidelberg), and scientists of the \emph{Herschel} Science Center (HSC). This paper makes use of the following ALMA data: ADS/JAO.ALMA\#2018.1.01003.S, ADS/JAO.ALMA\#2021.1.01365.S. ALMA is a partnership of ESO (representing its member states), NSF (USA) and NINS (Japan), together with NRC (Canada), NSTC and ASIAA (Taiwan), and KASI (Republic of Korea), in cooperation with the Republic of Chile. The Joint ALMA Observatory is operated by ESO, AUI/NRAO and NAOJ. This study was partly based on observations made with the NASA/DLR  SOFIA. SOFIA is jointly operated by the Universities Space Research Association Inc. (USRA), under NASA contract NNA17BF53C, and the Deutsches SOFIA Institut (DSI), under DLR contract 50 OK 0901 to the University of Stuttgart. upGREAT is a development by the MPIfR and the KOSMA/University Cologne, in cooperation with the DLR Institut f\"{u}r Optische Sensorsysteme. Financial support for FEEDBACK at the University of Maryland was provided by NASA through award SOF070077 issued by USRA. The FEEDBACK project is supported by the BMWI via DLR, project number 50 OR 2217 (FEEDBACK-plus).  The BLASTPol telescope was supported by through grant numbers NNX13AE50G, 80NSSC18K0481,
NAG5-12785, NAG5-13301, NNGO-6GI11G, NNX0-9AB98G, the Illinois Space Grant Consortium, the Canadian Space Agency, the Leverhulme Trust through the Research Project Grant F/00 407/BN, the Natural Sciences and Engineering Research Council of Canada, the Canada Foundation for Innovation, the Ontario Innovation Trust, and the US National Science Foundation Office of Polar Programs. L.M.F acknowledges support from the National Science and Engineering Research Council (NSERC) through Discovery Grant RGPIN/06266-2020, and funding through the Queen’s University Research Initiation Grant. G.N. is grateful for financial support for this work from NASA via award SOF06-0183 issued by USRA to Northwestern University. N.S. and R.S. acknowledge support by the SFB 1601, sub-project B2, funded by the DFG. T.G.S.P gratefully acknowledges support by the National Science Foundation under grant No. AST-2009842 and AST-2108989 and by NASA award \#09-0215 issued by USRA.

\facilities{SOFIA, ALMA, BLAST, APEX, Herschel, Spitzer, Mopra}
\software{Astropy \citep{astropy:2013, astropy:2018, astropy:2022}, NumPy \citep{harris2020array}, Matplotlib \citep{Hunter:2007}, Scipy \citep{2020SciPy-NMeth}, APLpy \citep{2012ascl.soft08017R}, Spectral-Cube \citep{2019zndo...2573901G}}

\begin{appendix}
\section{HRO Results using HAWC+ Band E}
\label{Results_BandE}

\begin{table}[htb!]
    \centering
    \caption{Same as Table \ref{tab:PRS} but using the magnetic field inferred from SOFIA/HAWC+ Band E (214 $\mu$m).}
    \begin{tabular}{ccccccc}
        \hline 
        \vspace{1mm}
        Instrument &
        Map & 
        Z$_{x}$ & $\sigma_{Z_{x}}$ & \Zprime & \Znorm & $n$  \\
        \hline \hline
        SPIRE & 500 $\mu$m  
        & -11.8 & 4.2 & -50.2 & -0.54 & 4350 \\
        SPIRE & 350 $\mu$m  
        & -12.2 & 3.9 & -48.2 & -0.52 & 4350 \\
        SPIRE & 250 $\mu$m  
        & -11.4 & 4.1 & -46.1 & -0.49 & 4350 \\
        HAWC+ & 214 $\mu$m 
        & -6.5 & 3.7 & -24.0 & -0.36 & 2225 \\
        PACS & 160 $\mu$m   
        & -6.8 & 3.6 & -24.5 & -0.35 & 4350 \\
        HAWC+ & 89 $\mu$m 
        & -3.3 & 7.8 & -25.8 & -0.22 & 6875 \\
        PACS & 70 $\mu$m   
         & -4.3 & 3.9 & -17.0 & -0.14 & 7252 \\
        IRAC & 4.5 $\mu$m   
        & +15.2 & 4.1 & +61.8 & 0.10 & 188273  \\
        IRAC & 3.6 $\mu$m   
        & +14.6 & 4.2 & +61.5 & 0.10 & 183000 \\
        ACA & 1.1-1.4mm  
        & -0.7 & 3.6 & -2.4 & -0.03 & 3342 \\
        12m & 1.1-1.4mm  
        & +0.1 & 3.9 & +0.5 & 0.00 & 39159  \\
        \emph{Herschel} & N(H$_{2}$) 
        & -9.4 & 3.8 & -35.9 & -0.38 & 4350 \\
        \emph{Herschel} & Temp
        & -1.9 & 3.9 & -7.3 & -0.08 & 4350 \\
        LAsMA & $^{12}$CO  
        & -1.5 & 4.1 & -5.9 & -0.06 & 4350 \\
        LAsMA & $^{13}$CO  
        & -8.0 & 4.1 & -32.9 & -0.35 & 4350 \\
        upGREAT & \CII\ 
        & -1.5 & 4.0 & -6.0 & -0.08 & 2799 \\
        upGREAT & \OI\   
        & -2.5 & 4.3 & -10.6 & -0.13 & 3209 \\
        Mopra & HNC  
         & -5.4 & 4.0 &  -22.0 & -0.24 & 4350 \\
        Mopra & C$^{18}$O 
         & -3.3 & 4.1 & -13.7 & -0.15 & 4350 \\
        Mopra & N$_{2}$H$^{+}$ 
         & -3.3 & 4.0 & -13.0 & -0.14 & 4350 \\
         \hline
    \end{tabular}
    \label{tab:PRS_BandE}
\end{table}

In this section, we present the HRO Single Map results (method described in Section \ref{sec:Methods_single}) using the HAWC+ Band E data to infer the magnetic field orientation. Table \ref{tab:PRS_BandE} gives the corrected and uncorrected PRS values for the different tracers. We find that the general trend of the Single Map results from the HAWC+ Band E data are fairly similar to the results found for Band C (see Table \ref{tab:PRS}). The consistency of the results is due to the similarity of the magnetic field morphologies traced in Band C and Band E (see Figure \ref{fig:BandC_BandE_HRO}).

\begin{figure}[thb!]
    \centering  
    \includegraphics[width=0.47\textwidth]{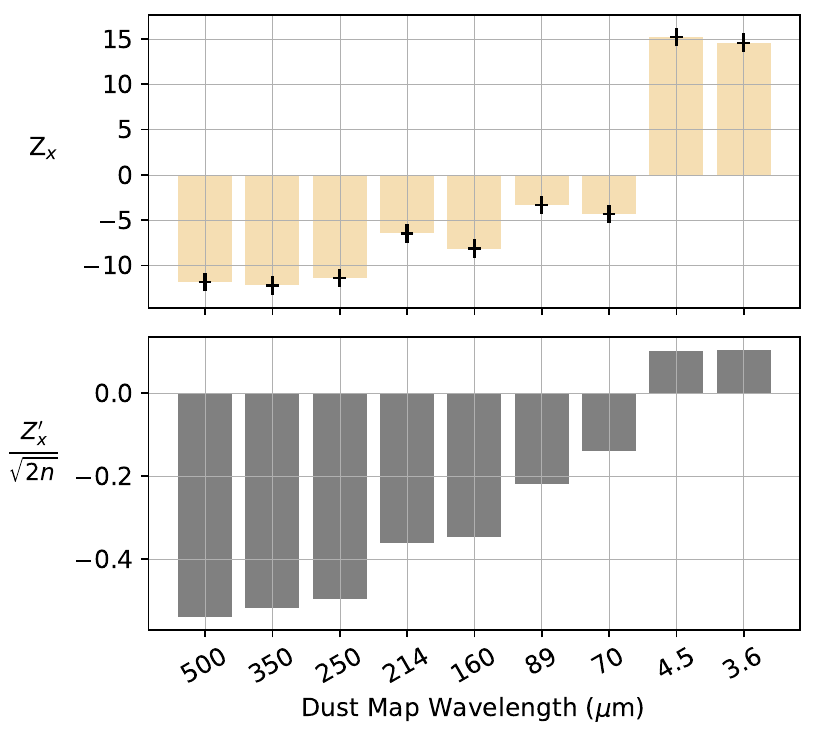}
    \caption{Same as Fig. \ref{fig:PRS_Wavelength} but showing HRO results using HAWC+ Band E (214 $\mu$m) polarization data to infer the magnetic field} 
    \label{fig:BandE_PRSWavelength}  
\end{figure}

Figure \ref{fig:BandE_PRSWavelength} shows the PRS (\Zx) for the different single-dish dust map wavelengths. In both Band C and Band E HRO analyses, the resulting sign (positive or negative) of \Zx\ as a function of dust wavelength is the same. The longer wavelength (500\textendash70 $\mu$m) dust maps show a statistically significant ($\lvert Z_{x} \rvert > 3$) preference for perpendicular alignment, while the \emph{Spitzer} maps show a statistically significant preference for parallel alignment. Similar to Band C, the Band E HRO results also show insignificant \Zx\ values for the ALMA continuum data. Furthermore, the Band E Single Map HRO results for the column density, temperature, and atomic and molecular lines are all consistent with the Band C results. Most gas tracers show a preference for perpendicular alignment, with the exception of \CII, $^{12}$CO, and \OI\ which have an overall statistically insignificant \Zx. The only difference between Band C and Band E results is the magnitude of the \Zx\ values. This is mostly due to the difference in resolution of the Band C and Band E data, since the magnitude of \Zx\ depends on the number of independent samples (as discussed in Section \ref{sec:Results_Single} and \ref{sec:PRS}).

Thus the Band E HRO results similarly find that tracers which are mostly sensitive to the N-S dense molecular ring and Main-Fil features show a preference for perpendicular alignment, while the tracers mostly sensitive to the E-W Bent-Fil features show a preference for parallel alignment. The interpretations of the results for Band C presented in Section \ref{sec:Interpret_HRO} are therefore also applicable to Band E.

\section{Uncertainty Estimation for the PRS}
\label{sec:PRS_error}

In Section \ref{sec:PRS}, we discussed how the oversampling-corrected Projected Rayleigh statistic \Zx\ is expected to have an uncertainty of 1 \citep{2018MNRAS.474.1018J}. These uncertainties do not, however, account for the measurement uncertainties in the magnetic field orientation \Bfield\ and the structure map $M(x,y)$, which may contribute additional sources of error in the HRO analysis. In this section, we perform Monte Carlo tests to propagate measurement uncertainties to the \Zx, \Zprime, and \Znorm\ calculations.

To estimate the impact of the uncertainties on the PRS due to \Bfield\ measurement uncertainties, we repeat the HRO analysis for 1000 magnetic field map iterations (\Bfield + \sigmaB), where a magnetic field orientation error term \sigmaB\ is added to the measured HAWC+ 89 $\mu$m (Band C) polarization angles. The error \sigmaB\ is drawn from a normal distribution centered at 0 with a standard deviation equal to the polarization angle error, which is estimated from the HAWC+ Data Reduction Pipeline (discussed in Section \ref{sec:SOFIA_reduction}). The uncertainty of the uncorrected statistic is then determined from the distribution of \Zprime\ values. We perform this test for two selected maps, the HAWC+ Band C intensity and the \CII\ integrated intensity map. We choose these maps since they have different alignment trends. The Band C intensity has a strong preference for perpendicular alignment ($Z_{x}=-9.3$), while \CII\ has  no clear statistical preference for parallel or perpendicular alignment ($Z_{x}=0.5$). For both maps, the Monte Carlo analysis with \Bfield\ uncertainties result in \Zprime\ and \Znorm\ distributions with standard deviations of $\sigma_{Z'_{x}}=0.2$ and $\sigma_{\Tilde{Z}'_{x}}=0.002$. The mean of the distributions is the same values as the Single Map HRO \Zprime\ and \Znorm\ values, which did not include measurement uncertainties (given in Table \ref{tab:PRS}).

To test whether measurement errors $\sigma_{M}$ in the intensity maps $M(x,y)$ could increase the uncertainty of our HRO analysis, we ran a Monte Carlo simulation which generates 1000 structure map iterations ($M(x,y) + \sigma_{M}$). We select the Band C intensity map for $M(x,y)$ since the uncertainties in total intensity have been estimated by the HAWC+ Data Reduction pipeline. We then smooth the maps by the Gaussian gradient kernel (using the method described in Section \ref{sec:ProjandMask} with the same kernel sizes listed in Table \ref{tab:Obs}) and calculate the relative angles for each iteration. We find that the standard deviation of the PRS values for this test to be to be $\sigma_{Z'_{x}}=0.8$ and $\sigma_{\Tilde{Z}'_{x}}=0.006$. We note that the uncertainties in the statistics are slightly larger in comparison to the polarization angle error propagation. However, neither the magnetic field orientation uncertainties nor the polarization angle uncertainties have a large impact on the final \Zx results. 

The Monte Carlo tests that have been presented in this section thus far have assumed that each pixel samples the probability distribution function of \Bfield\ independently from neighbouring pixels. However, since the FWHM beam area spans many pixels, the measurement errors are correlated between adjacent pixels. To estimate the uncertainties on the oversampling corrected \Zx, we re-calculate the PRS using only independent relative angle pixels (i.e one pixel per FWHM beam area). Using this approach, our Monte Carlo test with \Bfield\ measurement uncertainties (\sigmaB) gives a standard deviation of $\sigma_{Z_{x}}=0.2$ for both the Band C and \CII\ intensity maps, which is the same as the distribution in \Zprime\ found in the over-sampled case. The Monte Carlo test with $M(x,y)$ measurement uncertainties ($\sigma_{M}$) gives a standard deviation of $\sigma_{Z_{x}}=0.3$ for Band C intensity.  In all cases, \Zx\, \Zprime, and \Znorm\ have standard deviations less than 1. Therefore, the uncertainties in the PRS statistics are primarily due to the distribution in the relative orientation angles sampled at different locations in the map, rather than by measurement uncertainties in the maps or inferred magnetic field orientation. 

\section{Single Map HRO}
\label{sec:leftover_single}

\subsubsection{Dust Emission}
\label{sec:App_dust}

In this section, we show the remainder of the relative angle maps and histograms from the Band C (89 $\mu$m) Single Map HRO analysis that were not presented in Figure \ref{fig:Results_Summary}. 

\begin{figure*}
    \centering  
    \includegraphics[width=0.9\textwidth]{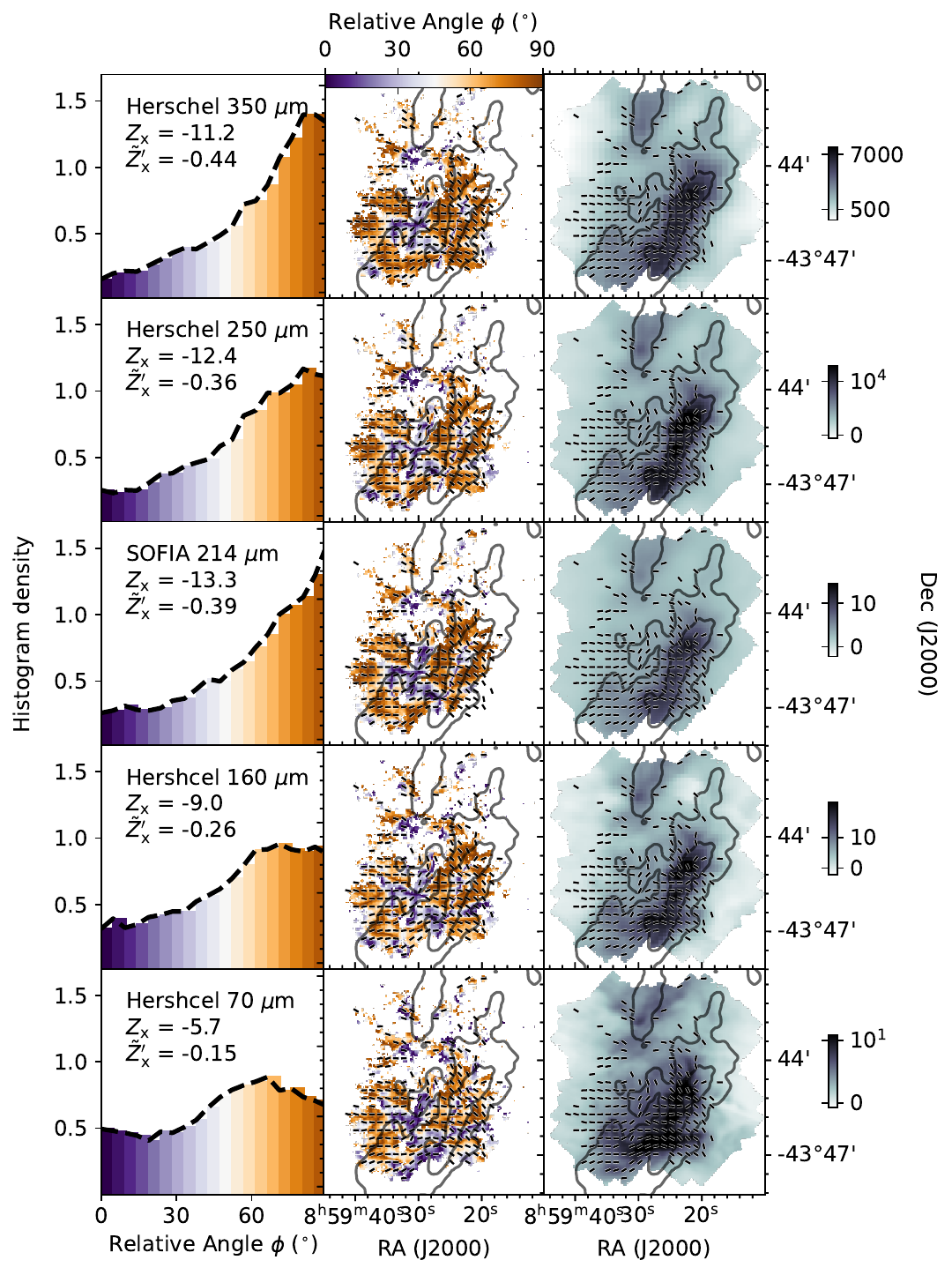}
    \caption{Same as Fig. \ref{fig:Results_Summary}. The right column is showing dust intensity with a log-scale colorbar in units of MJy sr$^{-1}$ for the 350 $\mu$m and 250 $\mu$m maps, and Jy pixel$^{-1}$ for the 214, 160 and 70 $\mu$m maps.} 
    \label{fig:Results_DustApp}  
\end{figure*}

Figure \ref{fig:Results_DustApp} shows the results for the 350--70 $\mu$m dust maps. We see that all maps trace the east and west halves of the N-S oriented dense ring which contribute most of the of the perpendicular $\phi(x,y)$ measurements. Comparing the different wavelengths maps, it can be seen that the emission from the longer wavelengths at 350 $\mu$m (first row) and 250 $\mu$m (second row) trace the ring structure most closely, particularly the denser western half which includes the Main-Fil, as outlined by the column density contours. This results in the higher degree of perpendicular alignment relative to the magnetic field, as signified by the higher magnitude \Znorm\ values for the longer wavelength dust maps in Table \ref{tab:PRS}.

One similarity for all dust maps at 350--70 $\mu$m in Fig. \ref{fig:Results_DustApp} is the alignment measured for the Flipped-Fil (labeled in Fig. \ref{figure:RGBScale}). All relative angle maps find $\phi\sim 0^{\circ}$ within the Flipped-Fil structure, even though the region is not always particularly noticeable in the intensity maps. This result of a mostly parallel relative alignment within the Flipped-Fil is consistent with the visual observation of the filament being elongated approximately in the direction of the N-S local magnetic field orientation, as seen in the middle and right panels of Fig. \ref{figure:RGBScale}. This is in contrast to the otherwise E-W orientation of the overall magnetic field morphology in the surrounding RCW 36 region.

The main difference between the different wavelengths is the emergence of the bright-rimmed Bent-Fil structures (labeled in Fig. \ref{figure:RGBScale}), particularly the south Bent-Fil which begins to appear over the south-western region of the dense ring at 160 $\mu$m and become increasing prominent at 70 $\mu$m. The Band C HAWC+ magnetic field lines are observed to largely follow the geometry of these Bent-Fils in the direction of their E-W elongation, resulting in increasingly parallel relative orientations at 160 $\mu$m and 70 $\mu$m, which decreases the overall magnitude of the \Zx\ to be less negative than at 350--214 $\mu$m. This trend can be also be seen in the left column HROs, which show a decreasing histogram density near $\phi \sim 90 ^{\circ}$ for the shorter wavelengths.

\begin{figure*}[h!]
    \centering  \includegraphics[width=0.9\textwidth]{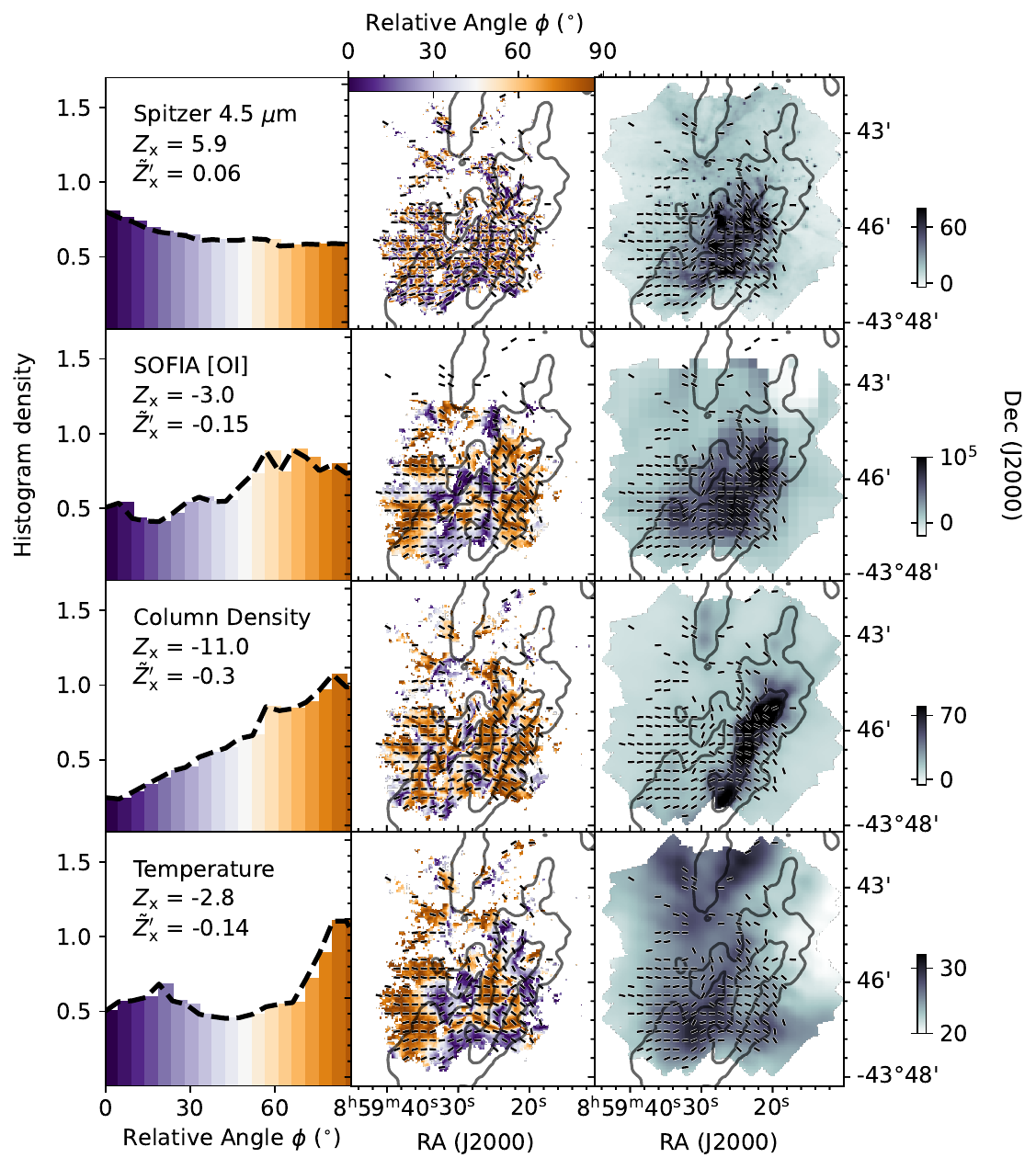}
    \caption{Same as Fig. \ref{fig:Results_Summary}. The right column colobar is in units of Jy pixel$^{-1}$ for the \textit{Spitzer} map, K km/s for the SOFIA \OI\ map, cm$^{-2}$ for the column density map and K for the temperature map. All colorbars are linear-scale.} 
    \label{fig:Results_Leftover}  
\end{figure*}

We contrast these results to the \emph{Spitzer} data at 4.5 $\mu$m, shown in the two row of Figure \ref{fig:Results_Leftover}. Similar to the 3.6 $\mu$m map shown in Fig. \ref{fig:Results_Summary}, the 4.5 $\mu$m map is also highly sensitive to the E-W Bent Fils structure, but does not trace the dense molecular ring. Since the Bent-Fils are oriented roughly parallel to the E-W magnetic field, this results in a statistically significant positive \Zx. In Section \ref{sec:Results_Single}, we made note of the apparent correlation between the emission of the Spitzer 3.6 $\mu$m map and the \CII\ total integrated intensity. Here, we once again note the similarities between the Spitzer emission and the \OI\ total integrated intensity map (shown in the second row of Fig. \ref{fig:Results_Leftover}), which is also a PDR tracer like \CII. The HRO results for the \OI\ data are further discussed in Section \ref{sec:App_Gas}. 

We now analyze the HRO results of the \textit{Herschel}-derived column density and temperature maps, shown in Figure \ref{fig:Results_Leftover}. Similar to longer wavelength dust maps, the column density map is also mostly sensitive to the dense molecular ring elongated in the roughly N-S oriented, particularly the western half containing the Main-Fil. This results in a majority of locally perpendicular $\phi(x,y)$ angles relative to the E-W magnetic field morphology, giving an overall large negative \Zx. In contrast, the temperature map shows only a slight preference for perpendicular relative alignment as it appears to be mostly tracing the bipolar morphology of the region. The HAWC+ magnetic field follows the curvature of the bipolar nebula and thus results in more parallel relative angles between the magnetic field and structures in the temperature map and a lower magnitude \Zx.

\begin{figure*}[thb!]
    \centering  
    \includegraphics[width = \textwidth]{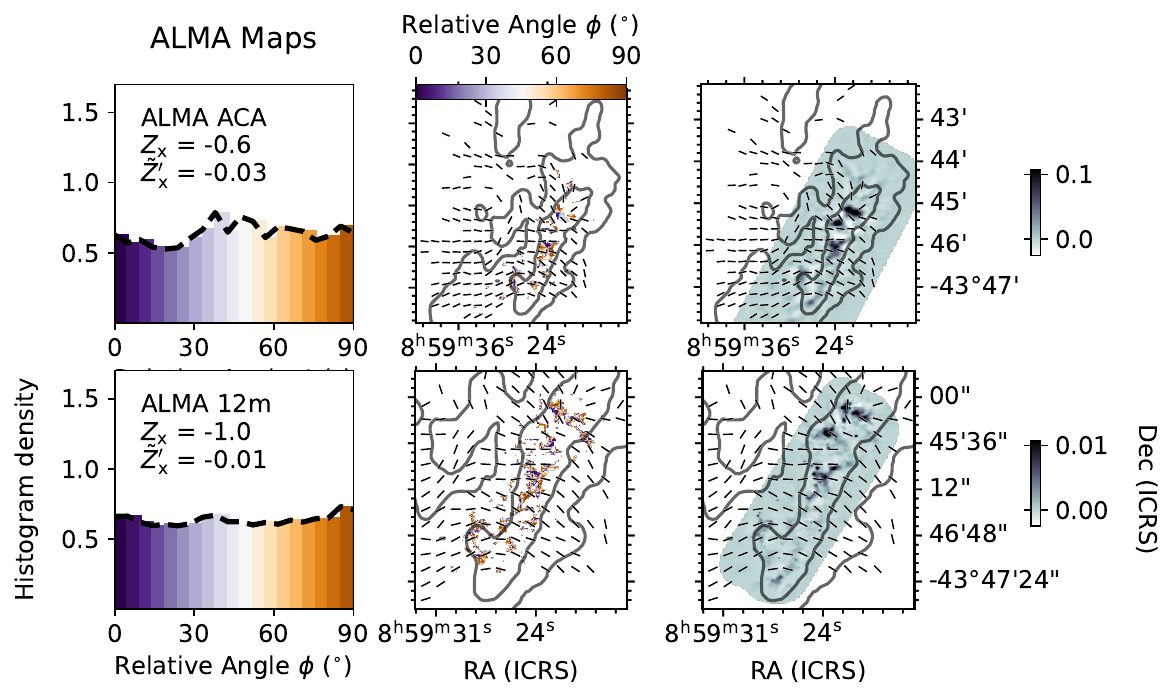}
    \caption{Same as Fig. \ref{fig:Results_Summary} but for ALMA data. The right column is showing a log-scale colorbar in Jy beam$^{-1}$ units for both continuum maps. Only ALMA data points above a 3-$\sigma$ signal-to-noise threshold are used for the HRO analysis.} 
    \label{fig:Results_ALMA}  
\end{figure*}

Figure \ref{fig:Results_ALMA} shows the results for the ALMA data. The HRO analysis for both the ALMA 12-m and ACA maps finds a statistically insignificant PRS of $Z_{x}\sim 0$, implying that the structures traced by ALMA do not have a preferred direction of orientation relative to the HAWC+ Band C inferred magnetic field. ALMA being an interferometer, resolves out much of the large scale structure such as the dense ring, Main-Fil, Flipped-Fil and Bent-Fils, which are the main features observed by the other dust maps. The HAWC+ Band C data is also likely not probing the magnetic field within the dense cores detected by ALMA as is further discussed in Appendix \ref{sec:CorrelateTracers}. This may explain why there is no correlation between the structures traced by the ALMA data and the magnetic field orientation inferred by SOFIA/HAWC+. Furthermore, there were not enough ALMA data points which were above a 3-$\sigma$ signal-to-noise threshold that also overlapped with the HAWC+ vectors, to produce a robust PRS measurement. An improved HRO study would compare the structures traced by the ALMA continuum maps to the magnetic field on similar core-scales, such as that inferred from ALMA polarization mosaics. This is outside the scope of this paper. 

\subsubsection{Spectral lines}
\label{sec:App_Gas}

\begin{figure*}[htb!]
    \centering  \includegraphics[width=0.9\textwidth]{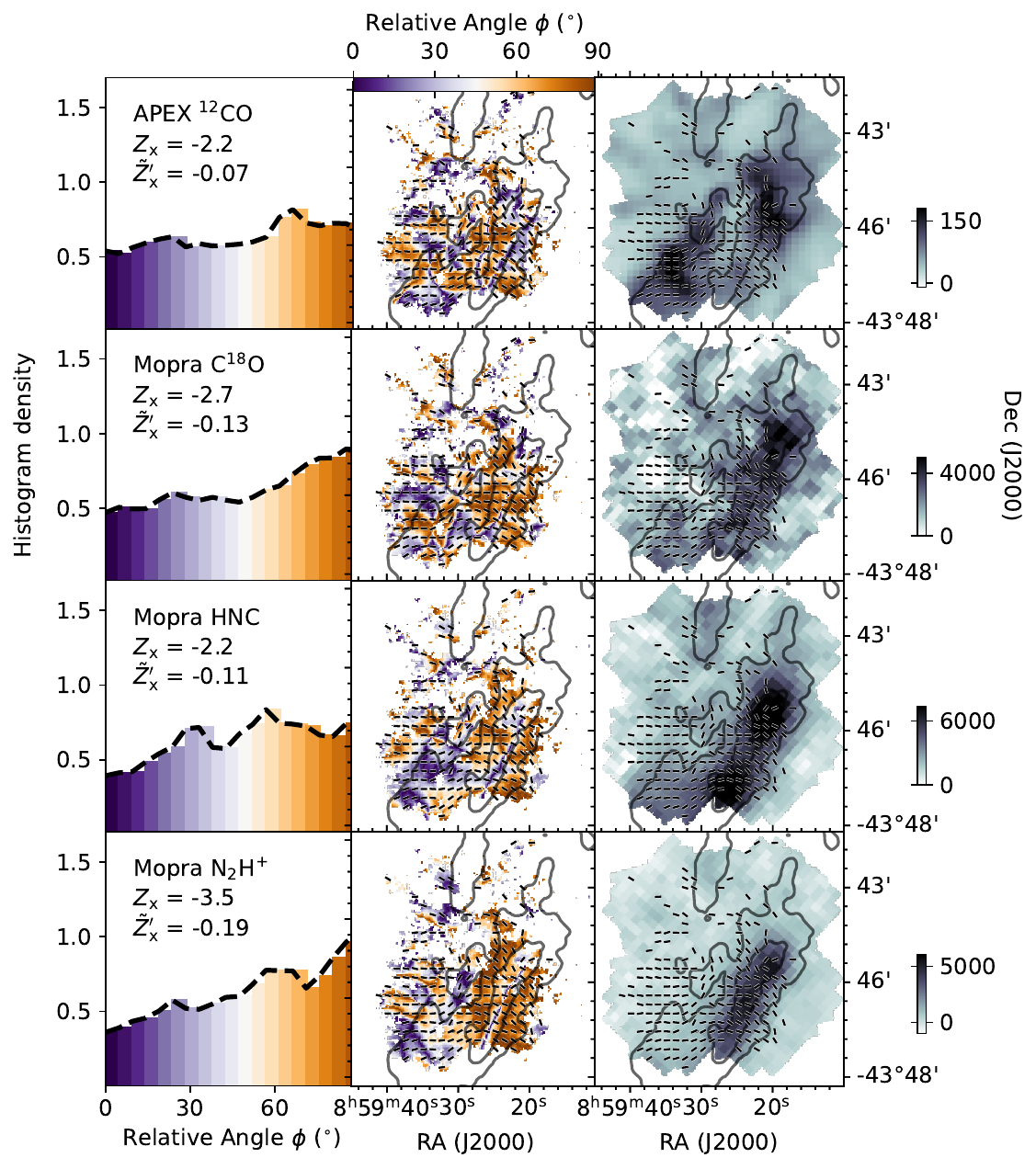}
    \caption{Same as Fig. \ref{fig:Results_Summary}. The right column is showing a integrated intensity in units of K km/s using a linear-scale colorbar for all maps.} 
    \label{fig:Results_GasApp}  
\end{figure*}

Figure \ref{fig:Results_GasApp} shows the Single Map HRO for the different molecular gas tracers. We compare these results with the atomic gas data of \OI\ in Figure \ref{fig:Results_Leftover} and \CII\ in Figure \ref{fig:Results_Summary}. For the atomic gas, the \CII\ and \OI\ data from SOFIA probe the transition from molecular to ionized gas in the PDR. The regions of parallel (purple) and perpendicular (orange) relative orientation angle in the \CII\ relative angle map is similar to \OI. The main difference between the two is that \CII\ \citep[with a critical density for collisional de-excitation of $\sim2.6\times 10^{3}$ cm$^{-3}$;][]{2006A&A...451..917R} traces more of the E-W oriented Bent-Fils in the west half of the ring as compared to the \OI. This may be because the Bent-Fils features in the western half are more diffuse, while \OI\ tends to trace denser PDR regions \citep[critical density of $5\times 10^{5}$ cm$^{-3}$;][]{2006A&A...451..917R} and hotter gas (typically $\sim$200 K). As such, \OI\ emission appears to better trace the N-S Main-Fil, resulting in a slight overall preference for perpendicular alignment as compared to \CII.

Next, we examine the Single Map HRO results for the molecular line data shown in Figure \ref{fig:Results_GasApp}. The high-density gas tracers such as HNC and N$_{2}$H$^{+}$ \citep[critical densities $> 10^{4}$ cm$^{-3}$;][]{2015PASP..127..299S, 2019ApJ...878..110F} clearly trace the densest N-S region in the west half of the ring where the alignment of the gas structures with respect to the magnetic field is mostly perpendicular within the Main-Fil column density contours, resulting in a negative \Zx\ value. The C$^{18}$O data  traces intermediate densities \citep[$\sim 10^{3}$ cm$^{-3}$;][]{2019ApJ...878..110F} similar to $^{13}$CO, but the C$^{18}$O Mopra data has a lower signal-to-noise ratio and lower resolution, resulting in a lower \Zx\ than the APEX $^{13}$CO data. Interestingly we find that the south Bent-fil traced by $^{13}$CO and HAWC+ 89 $\mu$m intensity (shown in Fig. \ref{fig:Results_Summary}), and can be somewhat seen in the maps of C$^{18}$O and HNC, but is not seen in the N$_{2}$H$^{+}$ map which tends to trace only high density and colder molecular gas. These observations are contrasted with the $^{12}$CO integrated intensity map which traces even lower density molecular gas and shows very different structure compared to the other spectral lines. While it also maps parts of the ring, $^{12}$CO shows bright emission around the Flipped-Fil and the north Bent-Fil which are elongated along the direction of the magnetic field lines resulting in an overall weak preference for perpendicular alignment relative to the magnetic field ($Z_x = -2.2$). Though $^{12}$CO is detected throughout the entire RCW 36 region, it is optically thick at the densest regions. Spectra of $^{12}$CO, $^{13}$CO, \CII\ and \OI\ for the Flipped-Fil are given in Appendix \ref{sec:leftover_velocity} for reference (for spectra in other regions, see Fig. 2 of \citealt{2022ApJ...935..171B}).

\subsubsection{Caveats and Considerations}
\label{sec:caveats}
Finally, we make note of some important considerations in our HRO analysis. We note that smoothing can reduce the number of data points near the boundaries of the map. For instance in the HAWC+ 89 $\mu$m map (shown in Fig. \ref{fig:Results_Summary}), the Gaussian kernel smoothing removes some of the relative angle measurements near the south edge of the map boundary which is not the case for the \emph{Herschel} 70 $\mu$m which covers a larger area on the sky (see Figure \ref{fig:Results_DustApp}). Another caveat to note for all our HRO analysis plots, is that some of the $\phi\sim0^{\circ}$ relative angle points are due to the gradient amplitude approaching zero as the gradient changes direction at the peak of the iso-contours. For example in Fig. \ref{fig:Results_Summary} at the center of the highest column density contour within the Main-Fil, a thin row of purple $\phi\sim0^{\circ}$ pixels can be seen along the N-S direction where we would expect the gradient to change direction. This is most obvious for the 350 $\mu$m and 250 $\mu$m relative angle maps. Since there is a small percentage of the total $\phi(x,y)$ pixels in the relative angle map which are subjected to this effect, the impact on the resulting \Zx\ is insignificant. Furthermore it should also be mentioned that in addition to the hot dust detected by \emph{Spitzer}, the instrument is also clearly detecting starlight from the massive stellar cluster. Since this emission is not extended but rather from point sources, it is generally not elongated in a particular direction and therefore also does not significantly affect \Zx.

\section{Velocity Dependent HRO}
\label{sec:leftover_velocity}

\begin{figure}[thb!]
    \centering  \includegraphics[width = 0.5\textwidth]{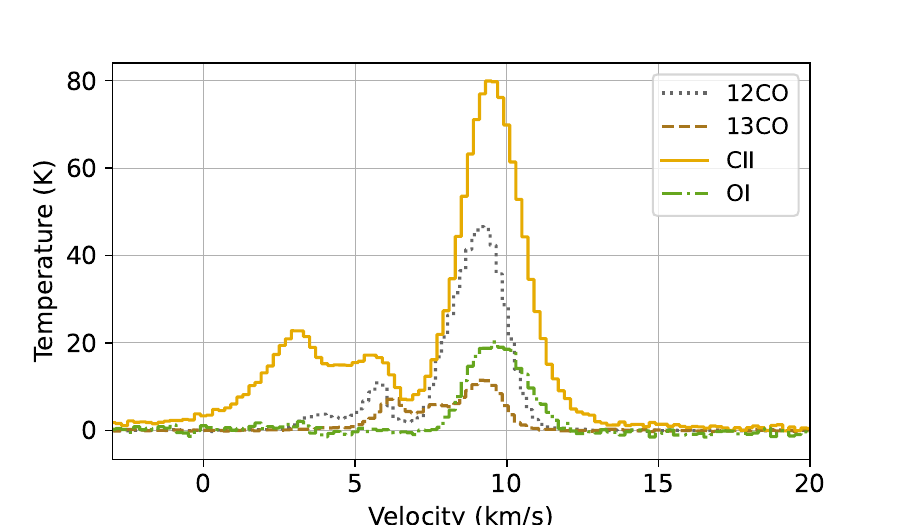}
    \caption{Spectra showing the antenna temperature of $^{12}$CO, $^{13}$CO, \CII\, \OI, for the Flipped-Fil region.} 
    \label{fig:Spectra}  
\end{figure}

\begin{figure*}[thb!]
    \centering  \includegraphics[width = 0.7\textwidth]{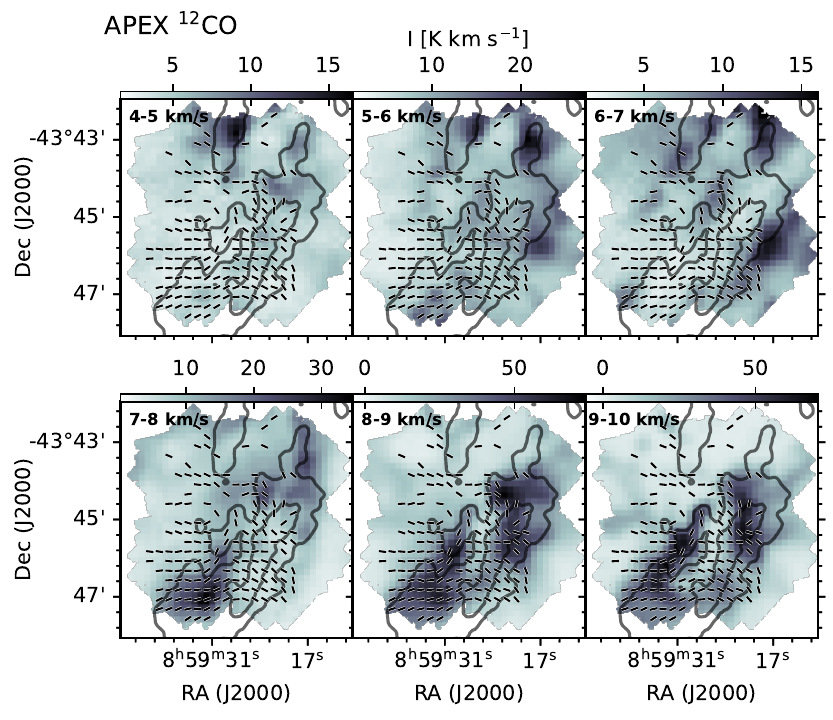}
    \caption{Same as Fig. \ref{fig:13CO_Slices}  
     but for APEX $^{12}$CO data.} 
    \label{fig:12CO_slices}  
\end{figure*}

\begin{figure*}[thb!]
    \centering  \includegraphics[width = 0.7\textwidth]{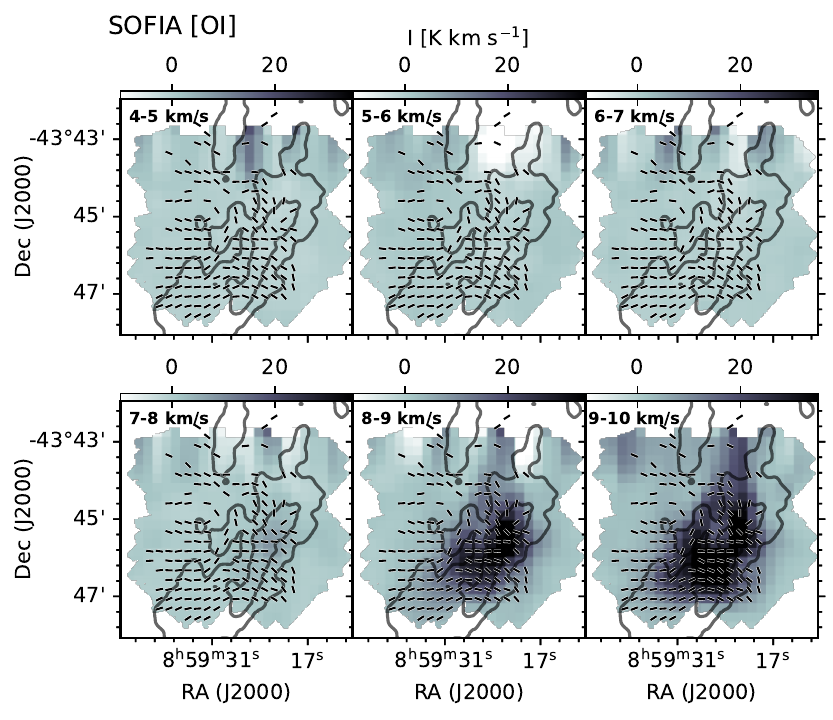}
    \caption{Same as Fig. \ref{fig:13CO_Slices}  
     but for SOFIA \OI\ data.} 
    \label{fig:OI_slices}  
\end{figure*}

\begin{figure*}[thb!]
    \centering  \includegraphics[width = 0.7\textwidth]{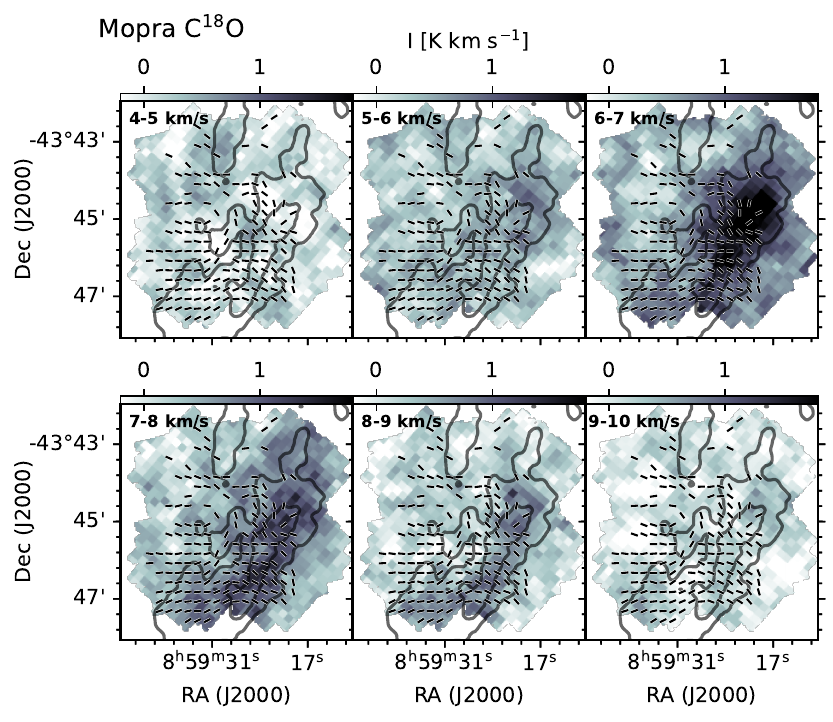}
    \caption{Same as Fig. \ref{fig:13CO_Slices}  
     but for Mopra C$^{18}$O data.} 
    \label{fig:C18O_slices}  
\end{figure*}

\begin{figure*}[thb!]
    \centering  \includegraphics[width = 0.7\textwidth]{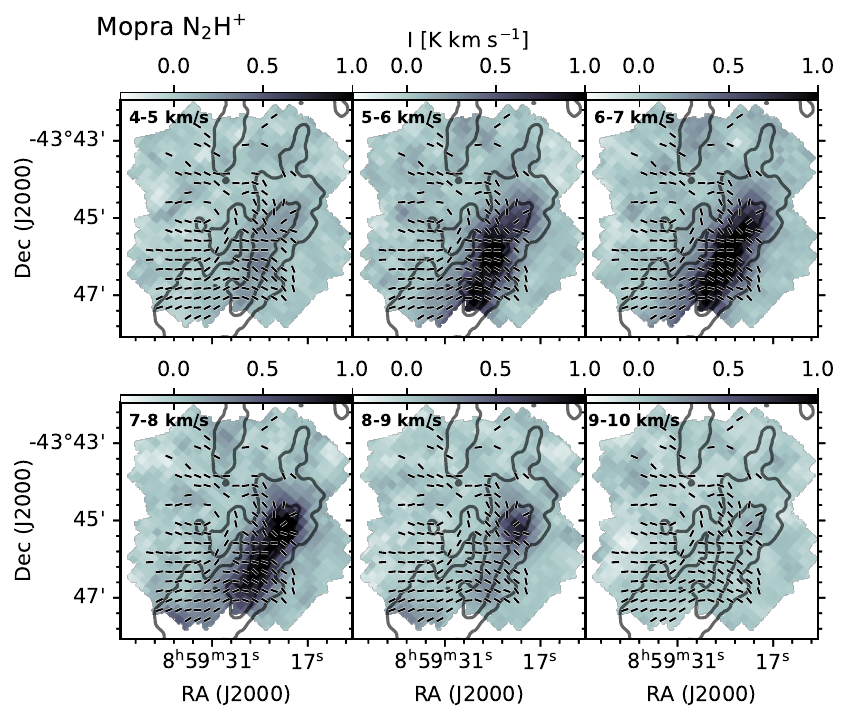}
    \caption{Same as Fig. \ref{fig:13CO_Slices}  
     but for Mopra N$_{2}$H$^{+}$ data.} 
    \label{fig:N2H_slices}  
\end{figure*}

In this section, we show the velocity channel maps for the gas tracers not included in Figures \ref{fig:13CO_Slices}\textendash\ref{fig:HNC_slices}. For reference, Figure \ref{fig:Spectra} shows the spectra for $^{12}$CO, $^{13}$CO, \CII\ and \OI\ at the Flipped-Fil region. Figures \ref{fig:12CO_slices}\textendash\ref{fig:N2H_slices} show the integrated intensities for 1 km/s wide velocity slabs from 4\textendash10 km/s for $^{12}$CO, \OI, C$^{18}$O, and N$_{2}$H$^{+}$, respectively. Similar to the main text, we note that the dense ring is traced at line-of-sight velocities of 4--7 km/s, while the Bent-Fils and Flipped-Fil are seen at 8--10 km/s.

\section{Optical Depth of HAWC+ Data}
\label{App:OptDept}

\begin{figure*}[thb!]
  \centering
  \includegraphics[width=\hsize]{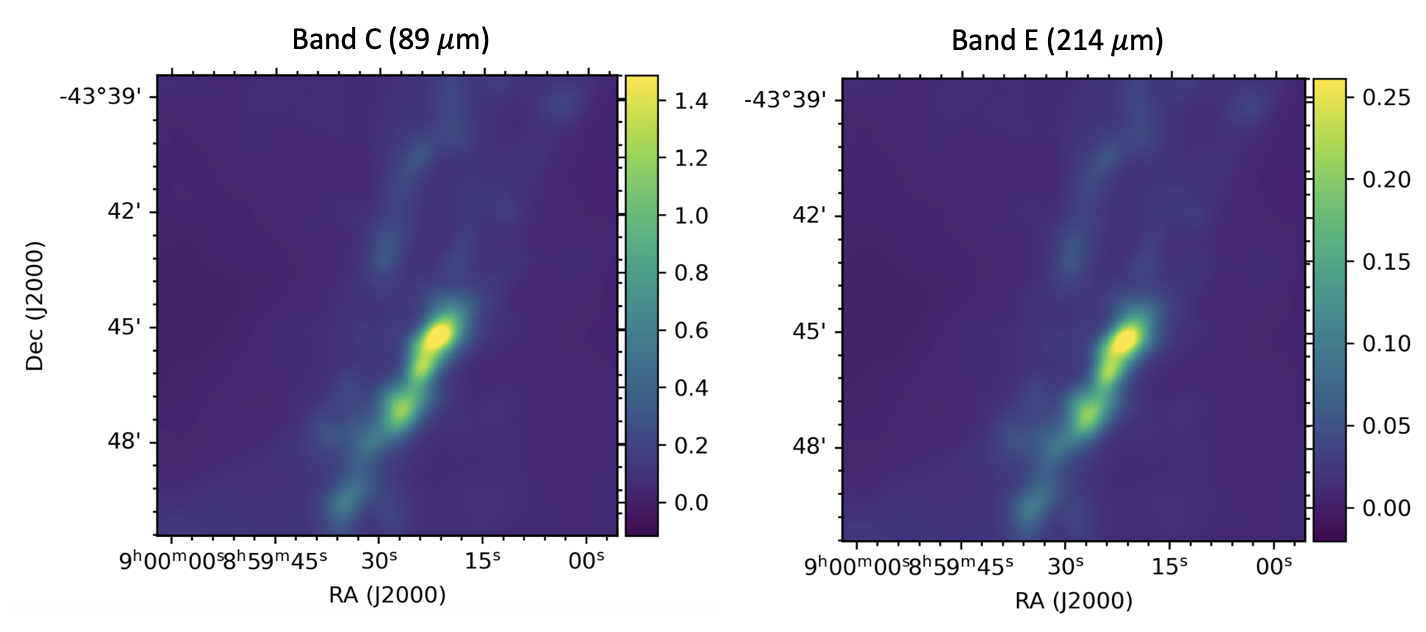}
  \caption{The optical depth $\tau$ (shown by the colorbar) of dust emission at 89 $\mu$m (\textit{left}) and 214 $\mu$m (\textit{right}), where $\tau <$1 corresponds to optically thin emission and $\tau >$ 1 corresponds to optically thick emission. The optical depth is estimated from \emph{Herschel}-derived column density and temperature maps at 36\arcsec{}.}
  \label{fig:OptDept}
\end{figure*}

Dust emission can be optically thin ($\tau \ll 1$), such that the plane-of-sky magnetic field orientation is inferred from the average emission by all dust grains along the line of sight. Alternatively, the emission can be optically thick ($\tau > 1$), such that only the flux from the outer surface layer of the cloud is traced. Understanding the optical depth helps identify the location of the dust grains dominating the emission, whether they are within a translucent cloud, or on the surface of an opaque cloud. To estimate the optical depth of RCW 36 at the HAWC+ wavelengths, 89 $\mu$m and 214 $\mu$m we use Equation \ref{eqn:tau} from the main text.

The \emph{Herschel}-derived column density (36\arcsec{} FWHM) map is used $N_{H_{2}}$. The 36\arcsec{} column density map is chosen over the 18\arcsec{} version as this resolution matches the temperature map. We use the dust opacity law from \cite{1983QJRAS..24..267H} that was applied by \cite{2011A&A...533A..94H} to originally derive the Vela C column density map, which is:

\begin{equation}
    \kappa_{\lambda} = 0.1 \times \left(\frac{\lambda}{300\,\mu m}\right)^{-\beta} 
\end{equation}

where we have used $\beta=2$, to match \cite{2011A&A...533A..94H}. The resulting estimates for optical depth $\tau$ are shown in Figure \ref{fig:OptDept}. From the colorbar, we see that at the longer HAWC+ wavelength of 214 $\mu$m (Band E), the dust emission is optically thin $\tau < 1$ everywhere in the region. At the shorter wavelength in HAWC+ at 89 $\mu$m (Band C), we estimate the emission is roughly optically thin everywhere, except at the brightest peaks near the clumps, where $\tau \sim 1.4$. While optically thick emission at an observing wavelength 89 $\mu$m would not be unexpected, in this case the maximum optical depth is still relatively close to the $\tau \sim 1$ surface, indicating that the emission is only moderately thick, rather than very thick (e.g for say $\tau \sim 10$). This means that we are missing some flux at the brightest peaks, but not too much.

However, we note that since we are using a 36\arcsec{} column density map to extrapolate the optical depth of the Band C emission, which has a native resolution of 8\arcsec{}, the actual optical depth at 8\arcsec{} could be higher than what is estimated from 36\arcsec{} FWHM map (shown in Fig. \ref{fig:OptDept}) at the smallest scales closest to the brightest peaks. Therefore, an optical depth of $\tau \sim 1.4$ at these points may be an underestimation. Another caveat is that a singular mass-weighted average dust temperature is assumed when generating the \emph{Herschel} column density map \citep{2014A&A...562A.138R}. This does not account for temperature gradients along the line of sight. If the 89 $\mu$m dust is probing a population of warmer dust grains, that only exist near the surface of the \HII\ region, then that dust will not be probing the entire column traced by the \emph{Herschel} column density map. Furthermore, the emission at 89 $\mu$m may also be tracing very small dust grains (VSGs) \citep{1968nim..book..221G}, which do not emit at sub-millimeter wavelengths but can at 70--100 $\mu$m \citep[e.g.,][]{1985ApJ...292..494D, 1987A&A...180...27W}.  These grains are stochastically heated and are not in equilibrium, which makes inferring their properties difficult. The 160--500 $\mu$m emission used to generate the column density map are likely tracing emission from the larger dust grains, meaning the column density map derived from these wavelengths will not include VSGs. 

\section{Correlation of Polarized Dust Emission with Other Tracers}
\label{sec:CorrelateTracers}

\begin{figure*}[htb!]
    \centering  \includegraphics[width=0.8\textwidth]{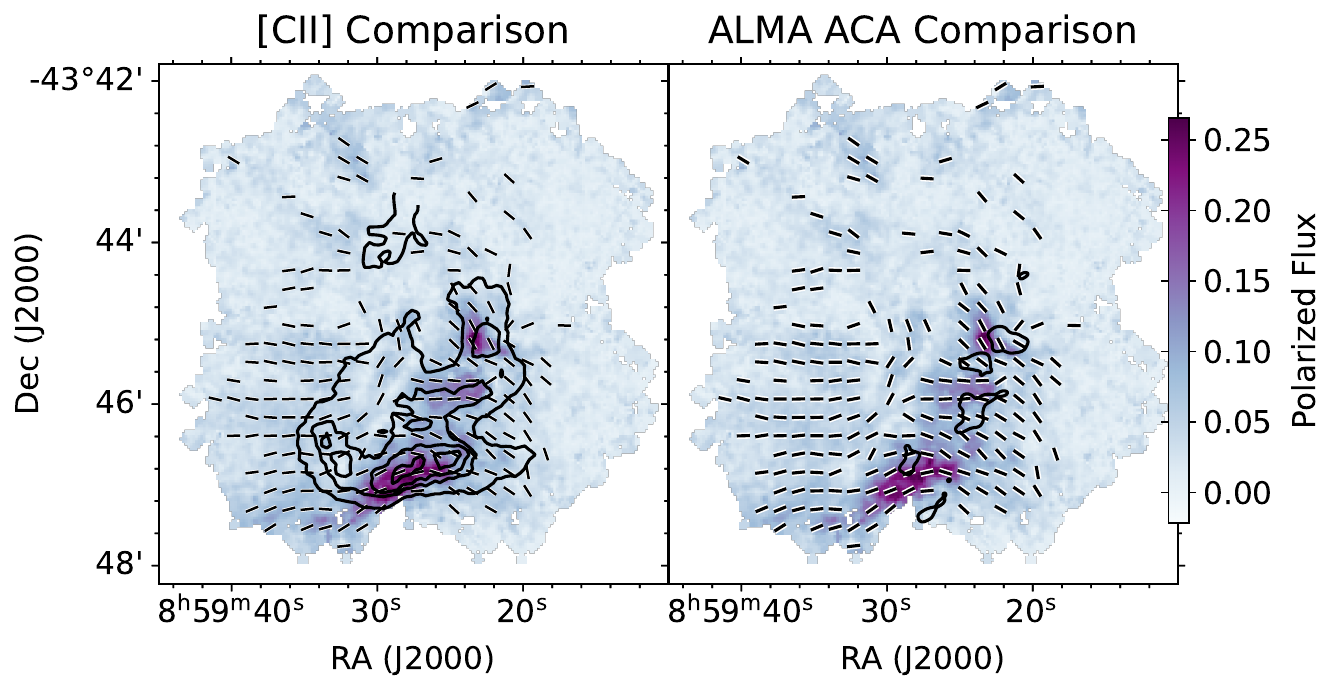}
    \caption{The colormap shows the polarized flux for HAWC+ Band C (89 $\mu$m) in both panels. \textit{Left:} shows the Band C polarized flux overlaid with SOFIA \CII\ integrated intensity contours at the levels of 300, 400, 450, 500 K km s$^{-1}$. \textit{Right:} shows the Band C polarized flux overlaid with ALMA ACA Band 6 1.1-1.4 mm continuum contours at the level of 0.018 Jy beam$^{-1}$.}
    \label{fig:Polarized_Overlay}  
\end{figure*}

In this section, we compare the 89 $\mu$m polarized emission to the other dust emission, column density, temperature and molecular line maps listed in Table \ref{tab:Obs}. These tracers probe different physical properties of the gas, and a strong correlation between the polarized emission and a particular dataset may imply the magnetic field is primarily being traced in regions with similar density, temperature, chemical and excitation conditions. 

We do this by individually overlaying contours of the different tracers on the Band C total polarization intensity map and visually comparing the emission.  Figure \ref{fig:Polarized_Overlay} shows that the 89 $\mu$m polarized emission correlates well with the \CII\ integrated intensity, shown in the left panel. This further reinforces our previous assertion that the polarization data is preferentially tracing the magnetic field from the warm dust located near the PDR, where [CII] is abundant. This is contrasted to the apparent anti-correlation of the Band C polarized data observed for the ALMA ACA continuum data, as can be seen in the right panel of Fig. \ref{fig:Polarized_Overlay}. The ALMA clumps appear to be located in areas where there is a lack of polarized intensity. This finding is consistent with Band C emission being sensitive to warmer dust, rather than the cold dense structures traced by ALMA. Polarization measurements at longer wavelengths and higher resolution (e.g. with ALMA) would be needed to probe the magnetic field within these colder dense structures.

\end{appendix}
\clearpage
\bibliography{references}{}
\bibliographystyle{aasjournal}
\end{document}